\documentclass[twocolumn,preprint]{aastex63}

\usepackage{amsmath,amstext}
\usepackage[T1]{fontenc}
\usepackage{apjfonts} 
\usepackage{natbib}
\usepackage{longtable}

\usepackage{microtype,longtable,verbatim,float,booktabs,graphicx}

\newcommand{\Ha}{\hbox{{\rm H}$\alpha$}}
\newcommand{\HavNII}{\hbox{({\rm H}$\alpha$+{\rm [N}\kern 0.1em{\sc II}{\rm ]})}}
\newcommand{\Hb}{\hbox{{\rm H}$\beta$}}
\newcommand{\SII}{\hbox{{\rm [S}\kern 0.1em{\sc ii}{\rm ]}}}
\newcommand{\NII}{\hbox{{\rm [N}\kern 0.1em{\sc ii}{\rm ]}}}
\newcommand{\OII}{\hbox{{\rm [O}\kern 0.1em{\sc II}{\rm ]}}}
\newcommand{\OIII}{\hbox{{\rm [O}\kern 0.1em{\sc iii}{\rm ]}}}
\newcommand{\NeIII}{\hbox{{\rm [Ne}\kern 0.1em{\sc III}{\rm ]}}} \newcommand{\NeIIIvHeI}{\hbox{{\rm [Ne}\kern 0.1em{\sc III}{\rm ]}+{\rm [He}\kern 0.1em{\sc I}{\rm ]}}}
\newcommand{\NeV}{\hbox{{\rm [Ne}\kern 0.1em{\sc v}{\rm ]}}}
\newcommand{\HeII}{\hbox{{\rm He}\kern 0.1em{\sc II}}}
\newcommand{\HII}{\hbox{{\rm H}\kern 0.1em{\sc II}}}
\newcommand{\Ms}{M_\star}

\begin{document}

\title{\large \bf CLEAR: Emission Line Ratios at Cosmic High Noon}

\author[0000-0001-8534-7502]{Bren E. Backhaus}
\affil{Department of Physics, University of Connecticut, Storrs, CT 06269, USA}

\author[0000-0002-1410-0470]{Jonathan R. Trump}
\affil{Department of Physics, University of Connecticut, Storrs, CT 06269, USA}

\author[0000-0001-7151-009X]{Nikko J. Cleri}
\affil{Department of Physics, University of Connecticut, Storrs, CT 06269, USA}
\affiliation{Department of Physics and Astronomy, Texas A\&M University, College
Station, TX, 77843-4242 USA}
\affiliation{George P.\ and Cynthia Woods Mitchell Institute for
Fundamental Physics and Astronomy, Texas A\&M University, College
Station, TX, 77843-4242 USA}

\author[0000-0002-6386-7299]{Raymond Simons}
\affil{Space Telescope Science Institute, 3700 San Martin Drive,
Baltimore, MD, 21218 USA}

\author[0000-0003-1665-2073]{Ivelina Momcheva}
\affil{Space Telescope Science Institute, 3700 San Martin Drive,
Baltimore, MD, 21218 USA}

\author[0000-0001-7503-8482]{Casey Papovich}
\affiliation{Department of Physics and Astronomy, Texas A\&M University, College
Station, TX, 77843-4242 USA}
\affiliation{George P.\ and Cynthia Woods Mitchell Institute for
Fundamental Physics and Astronomy, Texas A\&M University, College
Station, TX, 77843-4242 USA}

\author[0000-0001-8489-2349]{Vicente Estrada-Carpenter}
\affiliation{Department of Physics and Astronomy, Texas A\&M University, College
Station, TX, 77843-4242 USA}
\affiliation{George P.\ and Cynthia Woods Mitchell Institute for
Fundamental Physics and Astronomy, Texas A\&M University, College
Station, TX, 77843-4242 USA}

\author[0000-0001-8519-1130]{Steven L. Finkelstein}
\affil{Department of Astronomy, The University of Texas, Austin, Texas, 78712 USA}

\author[0000-0002-7547-3385]{Jasleen Matharu}
\affiliation{Department of Physics and Astronomy, Texas A\&M University, College
Station, TX, 77843-4242 USA}
\affiliation{George P.\ and Cynthia Woods Mitchell Institute for
Fundamental Physics and Astronomy, Texas A\&M University, College
Station, TX, 77843-4242 USA}

\author[0000-0001-7673-2257]{Zhiyuan Ji}
\affil{Department of Astronomy, University of Massachusetts,
Amherst, MA, 01003 USA}

\author[0000-0001-6065-7483]{Benjamin Weiner}
\affil{MMT/Steward Observatory, 933 N. Cherry St., University of Arizona, Tucson,
AZ 85721, USA}

\author{Mauro Giavalisco}
\affil{Department of Astronomy, University of Massachusetts,
Amherst, MA, 01003 USA} 

\author[0000-0003-1187-4240]{Intae Jung}
\affil{Department of Physics, The Catholic University of America, Washington, DC 20064, USA}
\affil{Astrophysics Science Division, Goddard Space Flight Center, Greenbelt, MD 20771, USA}

\begin{abstract}
We use \textit{Hubble Space Telescope} WFC3 G102 and G141 grism spectroscopy to measure rest-optical emission-line ratios of 533 galaxies at $z\sim1.5$ in the CANDELS Ly$\alpha$ Emission at Reionization (CLEAR) survey. We compare $\OIII/\Hb$ vs. $\SII/\HavNII$ as an ``unVO87'' diagram for 461 galaxies and $\OIII/\Hb$ vs. $\NeIII/\OII$ as an ``OHNO'' diagram for 91 galaxies. The unVO87 diagram does not effectively separate active galactic nuclei (AGN) and $\NeV$ sources from star-forming galaxies, indicating that the unVO87 properties of star-forming galaxies evolve with redshift and overlap with AGN emission-line signatures at $z>1$. The OHNO diagram does effectively separate X-ray AGN and $\NeV$-emitting galaxies from the rest of the population.
We find that the $\OIII/\Hb$ line ratios are significantly anti-correlated with stellar mass and significantly correlated with $\log(L_{\Hb})$, while $\SII/\HavNII$ is significantly anti-correlated with $\log(L_{\Hb})$. Comparison with MAPPINGS~V photoionization models indicates that these trends are consistent with lower metallicity and higher ionization in low-mass and high-SFR galaxies. We do not find evidence for redshift evolution of the emission-line ratios outside of the correlations with mass and SFR.
Our results suggest that the OHNO diagram of $\OIII/\Hb$ vs. $\NeIII/\OII$ will be a useful indicator of AGN content and gas conditions in very high-redshift galaxies to be observed by the \textit{James Webb Space Telescope}.

\end{abstract}


\section{Introduction}

Emission lines from galaxies can be used to determine their physical properties and evolution history, including their star formation, metallicity, ionization, and other gas conditions. One classic way to present emission-line data is by comparing ratios of lines, the most common of which are the BPT \citep{bald81} and VO87 \citep{veil87} diagrams. The BPT and VO87 diagrams compare the emission-line ratios $(\NII/\Ha)$ to $(\OIII/\Hb)$ and $(\SII/\Ha)$ to $(\OIII/\Hb)$, respectively, as these are some of the strongest emission lines in rest-frame optical spectra and the close wavelengths of each pair lessens the effect of dust attenuation. The BPT and VO87 diagrams can be used
to classify galaxies as star forming (SF), active galactic nuclei (AGN), or low-ionization narrow emission-line regions (LINERs), by using the emission-line ratios to infer the relative ionization of the gas \citep{bald81,veil87,kauf03,kewl06}. 

Recent advances in near-infrared spectrographs have extended studies of rest-frame optical emission lines to galaxies at $z>1$. This redshift range includes the peak of cosmic star formation \citep{mada14} and supermassive black hole growth \citep{aird10}, therefore it is particularly interesting to understand the physical conditions of galaxies during the era that includes most of their assembly and evolution. 

Observations of $z>1$ galaxies show systematic offsets in emission-line properties compared to low-redshift galaxies, with high-redshift galaxies exhibiting higher ratios of high-ionization metal lines to Balmer lines. These emission-line observations have been interpreted using a diverse range of theoretical explanations. 
The "R23", (\OIII $\lambda$ 4959,5007 +\OII $\lambda$ 3726,3729)/\Hb, emission-line ratio demonstrates that high-redshift galaxies have lower metallicity, largely due to typically having lower stellar masses
\citep{kobu04, zahi13, henr13, wuyt14}.
In addition, ratios of partially ionized metal lines with Balmer lines indicate higher ionization \citep{Stei14,shap15,strom18}, show evidence for higher electron density \citep{Brin08,Liu08,sand20,Runc21} and higher gas pressure \citep{Kewl19} for high-redshift galaxies.
High-redshift galaxies also have unusual N/O abundance patterns, which
may be due to different $\alpha$-capture fusion and/or Wolf-Rayet stars \citep{mast14}.
Spatially resolved emission line gradients \citep{Trump11,Trump14} and
X-ray observations \citep{xue12} additionally suggest a larger presence of AGN in $z>1$ galaxies.

\begin{deluxetable}{l|r}[t]
\tablecaption{Definitions \label{tab:mlinmix}}
\tablenum{1}
\tablecolumns{2}
\tablewidth{0pt}
\tablehead{\colhead{Name} & \colhead{Lines}}
\startdata
VO78 & $\OIII/\Hb$ vs. $\SII/\Ha$\\
unVO87 & $\OIII/\Hb$ vs. $\SII/\HavNII$ \\
OHNO &  $\OIII/\Hb$ vs. $\NeIII/\OII$\\
\enddata
\end{deluxetable}

The near-infrared sky background strongly limits emission-line observations of high-redshift galaxies from the ground. Bright OH lines and water vapor absorption often limit studies to single pairs of emission lines, which tend to poorly constrain the physical conditions and AGN content of high-redshift galaxies \citep[e.g.,][]{Trump11,coil15}. Ground-based studies that compare different emission-line pairs in the BPT and VO87 diagrams are typically limited to small samples of bright and/or massive galaxies within narrow redshift ranges.

\begin{figure*}[t]
\epsscale{1.1}
\plotone{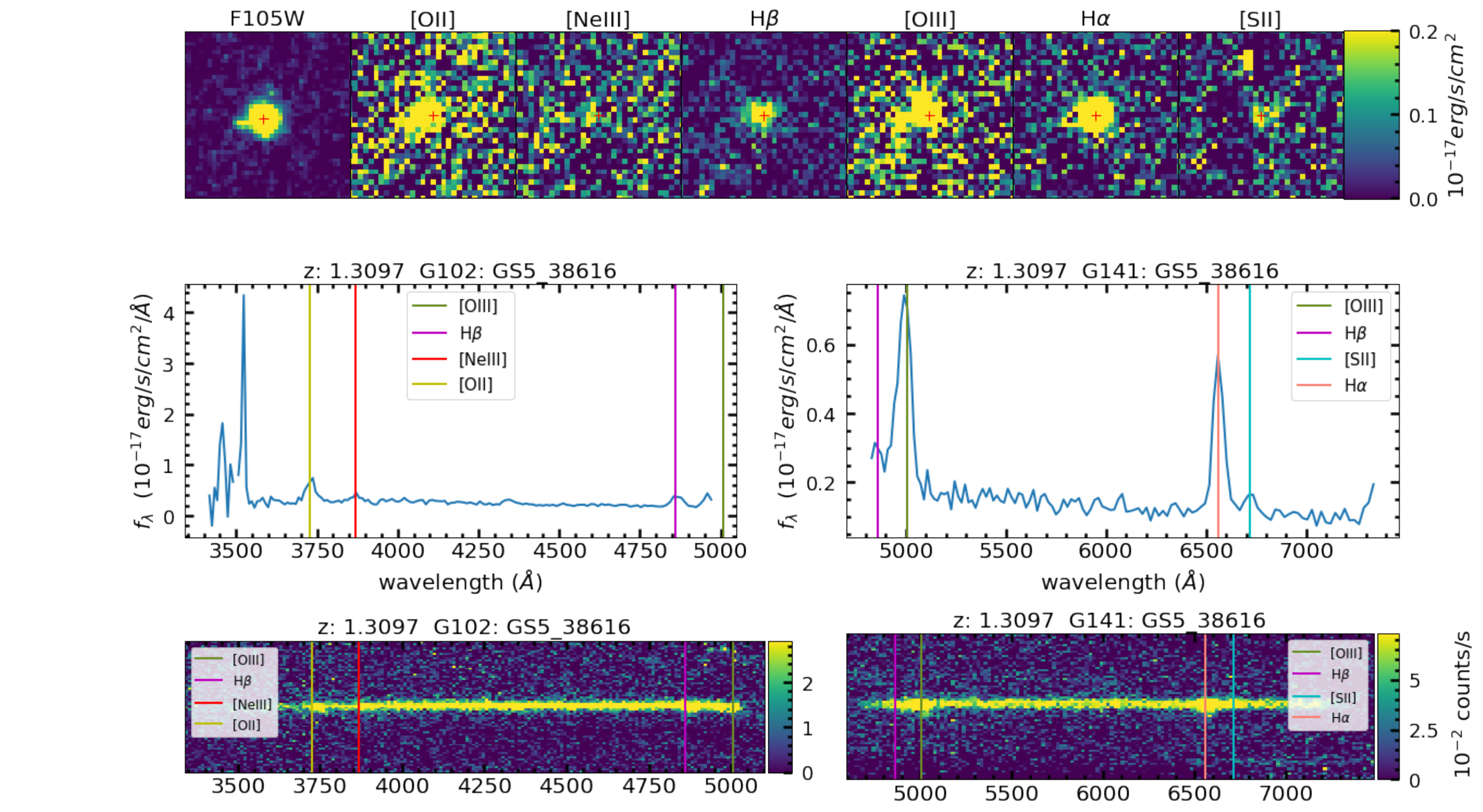}
\caption{The emission-line maps (\textit{top}), one-dimensional (1D) (\textit{middle}),  and two-dimensional (2D) G102 and G141 spectra (\textit{bottom}) for an example galaxy, GS-38616, that is in both of our unVO87 and OHNO samples. The red cross in the images denotes the center of the galaxy. Vertical lines in the 1D and 2D spectra indicate the emission lines of the unVO87 and OHNO diagrams
\label{fig:spec}}
\end{figure*}
 
In this work we use near-infrared spectroscopy from the \textit{Hubble Space Telescope} (\textit{HST}) Wide Field Camera 3 (WFC3) slitless grisms, taken as part of the CANDELS Ly$\alpha$ Emission at Reionization (CLEAR) \citep{estr19} and 3D-HST \citep{momc16} surveys, to investigate the rest-frame optical emission-line properties of 533 galaxies at $0.6<z<2.5$. The combination of the G102 and G141 slitless grisms gives full observed-frame wavelength coverage over $0.8<\lambda<1.7~\mu$m, and with less sky background than ground-based observations.

\begin{figure*}[tbp]
\centering
\epsscale{0.9}
\plotone{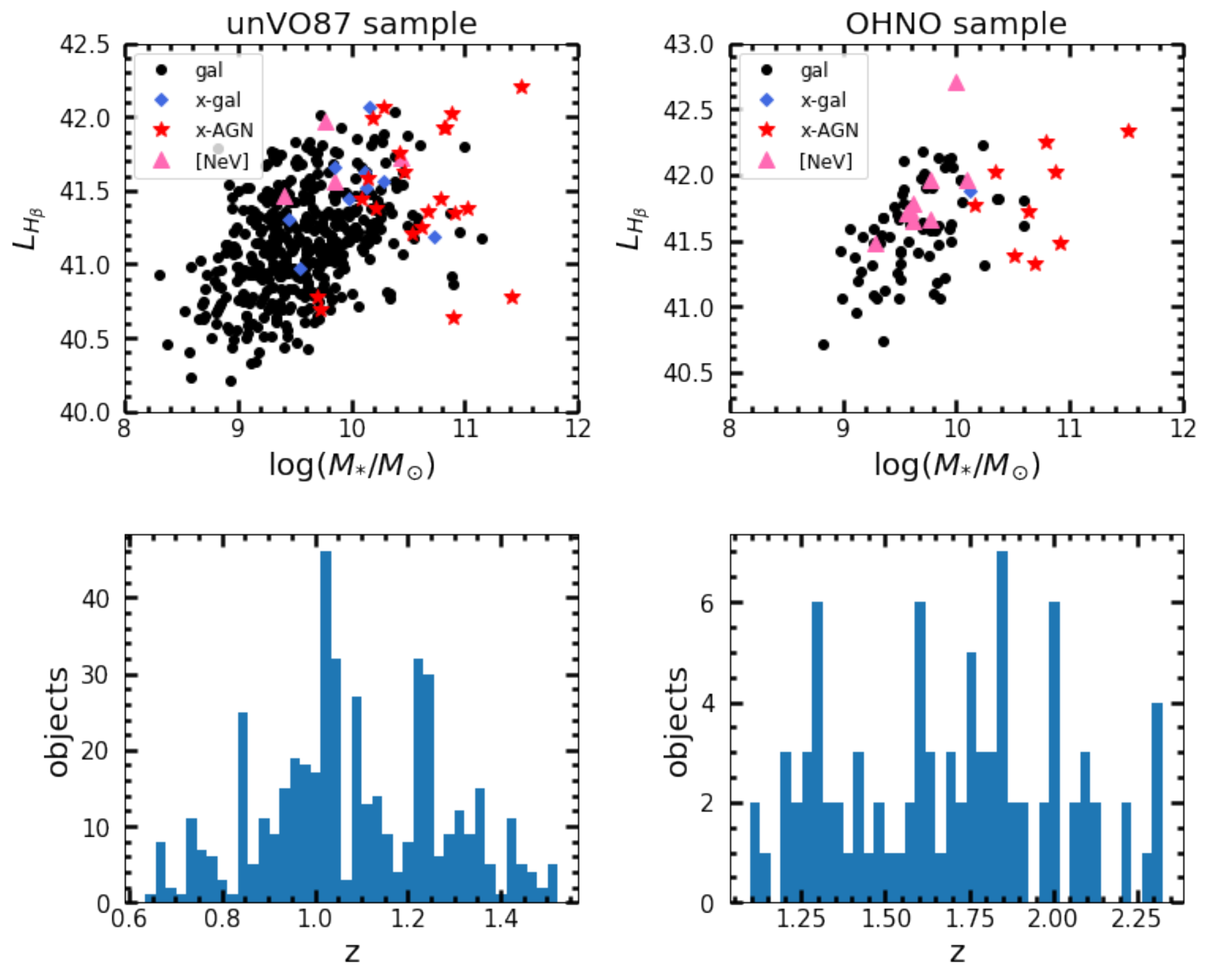}
\caption{The distribution of \Hb\ luminosity and stellar mass for the unVO87 (top left) and OHNO (top right) galaxy samples. The bottom panels show the redshift distributions of the two samples. We use the \Hb\ luminosity as a proxy for star formation rate, following Equation \ref{Eq:SFR}. Regular galaxies, X-ray galaxies, X-ray AGN and \NeV\ sources are marked by black points, blue diamonds, red stars and pink triangles respectively in the top two panels.
\label{fig:sample}} 
\end{figure*} 

\begin{figure*}[tbp]
\centering
\epsscale{0.9}
\plotone{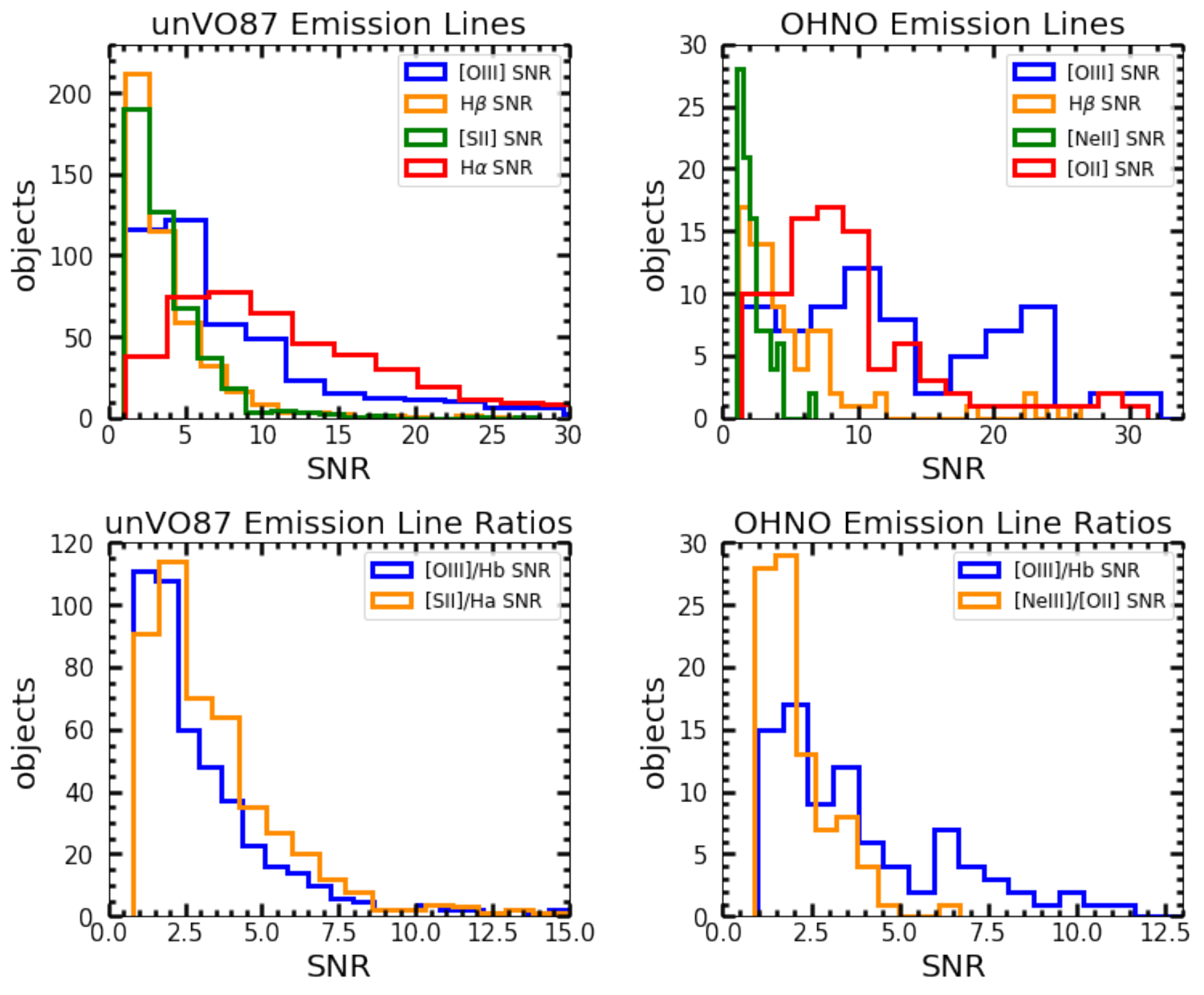}
\caption{The distribution of emission line SNR of the galaxies in our unVO87 (top left) and OHNO (top right) samples. The bottom panels show the SNR distributions of the emission-line ratios of the two samples. The vast majority (99\%) of galaxies in both samples have at least one emission line detected at 3$\sigma$, such that even a galaxy has SNR$<$3 detections of weaker lines it still has well-constrained emission-line ratios.
\label{fig:SNR}} 
\end{figure*} 

We describe the \textit{HST}/WFC3 grism observations and introduce the sample in Section 2. In Section 3 we introduce and discuss the OHNO and unVO87 diagrams.  We compare our high-redshift CLEAR sample to low-redshift ($z\sim0.06$) Sloan Digital Sky Survey (SDSS) galaxies in Section 4, and discuss the evolution of the OHNO and unVO87 diagrams through time. Finally, Section 5 describes the connections between galaxy properties and gas conditions inferred from the emission-line ratios. Throughout this work, we assume a $\Lambda$CDM cosmology with $\Omega_{M}=0.3$, $\Omega_{\Lambda}=0.7$ and $H_{0}=70$ km $s^{-1}$ $Mpc^{-1}$ \citep{plan15}.

\section{Observational Data and Sample}

\subsection{HST Spectroscopy}

Our galaxy sample comes from \textit{HST} near-IR spectroscopy with the G102 and G141 grisms taken as part of the 3D-HST \citep{momc16,bram12,vand11} and CLEAR surveys \citep{estr19}. The G102 observations come from programs GO-14227 (PI Papovich), GO-13420 (PI Barro), GO/DD-11359 (‘ERS’, PI O’Connell), and GO-13779 (‘FIGS’, PI Malhotra). The G141 observations come from programs GO-11600 (‘AGHAST’; PI Weiner), GO-12461 (‘SN COLFAX’, PI Reiss), GO-13871 (PI Oesch), GO/DD-11359 (‘ERS’, PI: O’Connell), GO12099 (‘GEORGE, PRIMO’, PI Reiss), and GO-12177 (‘3D-HST’, PI van Dokkum). CLEAR includes WFC3 G102 slitless grism spectroscopy that covers $0.8-1.15$~\micron\ in 12 fields between GOODS-North (GN) and GOODS-South (GS) to a 12-orbit depth (\citealt{estr19, Simo20}). The CLEAR pointings overlap with the 3D-HST survey \citep{momc16}, which gives G141 slitless grism spectra covering $1.1-1.65$~\micron\ with a 2-orbit depth. These combined datasets provide low-resolution grism spectroscopy over observed-frame $0.8<\lambda<1.65$~\micron\ for every source in the field of view, covering a suite of rest-frame optical lines in a large number of $0.6<z<2.5$ galaxies. The grism spectra have low \textit{spectral} resolution, with $R \sim 210$ in G102 and $R \sim 130$ in G141 for point sources. In contrast, the two-dimensional spectra have high \textit{spatial} resolution of $0\farcs06$ per pixel.

The grism data were reduced using the \texttt{grizli} (grism redshift and line analysis) software\footnote{\texttt{https://github.com/gbrammer/grizli/}}, as described in \citealt{Simo20}. 
The high spatial resolution and low spectral resolution of the slitless grism observations mean that features in the two-dimensional spectra are generally caused by the spatial morphology of the source rather than by kinematics. The \texttt{grizli} software directly fits the 2D spectra using model spectra convolved with the galaxy image, while subtracting contamination from overlapping spectra using the model fits for nearby objects. Spectra observed at different position angles are fit independently. We use spectroscopic redshifts and line fluxes from the CLEAR catalog v3.0.0 \citep{Simo20}. An example of the \texttt{grizli}-reduced 1D and 2D spectra is shown in Figure \ref{fig:spec}.

We do not dust correct the emission lines of our CLEAR sample because the \Ha\ emission line is not available for both samples: the low grism resolution blends the \Ha\ with \NII, and much of the OHNO galaxy sample is at too high redshift to have \Ha\ in the observed-frame spectra. The unVO87 and OHNO diagrams themselves are not effected by dust attenuation because the emission line pairs are close in wavelength and are nearly equally effected by dust. Dust attenuation will only effect our analysis when using emission line luminosities in Section 4, and in this section we discuss how dust attenuation effects our interpretation.

Stellar masses for objects in our sample are calculated following the approach in the 3D-HST catalog \citep{skel14}, recalculated from the CANDELS imaging \citep{grog11, koek11} with added CLEAR F105W photometry. The stellar masses are calculated using FAST \citep{krie09}, with the \cite{bruz03} stellar population synthesis model library, a \cite{chab03} IMF, solar metallicity, and assuming exponentially declining star formation histories. 


We also used the X-ray classifications from \cite{xue16} for the GN fields and \cite{luo17} for the GS fields. The GS X-ray catalogue used a 7Ms exposure covering 484.2 arcmin$^2$ while the GOODS-N X-ray catalogue used a 2Ms exposure. X-ray sources are classified as ``AGN'' if they have luminosities above $10^{42}$~erg~s$^{-1}$ or X-ray to optical flux ratios of $\log(f_x/f_R)>-1$ \citep{luo17, xue16}. The GOODS-S catalog \citep{luo17} additionally used X-ray to infrared flux ratios, X-ray to radio flux ratios, and spectroscopy (with broad emission lines) to classify X-ray sources as AGN.

\subsection{Sample Selection}

We create two samples of galaxies based on $\mathrm{SNR}>1$ detection of sets of four (integrated) emission lines. The ``unVO87'' sample includes galaxies with $\mathrm{SNR}>1$ detections of the \OIII$\lambda$5007, \Hb, blended \SII$\lambda$6718+6732 doublet, and blended \Ha+\NII$\lambda$6584 emission lines. This sample is analogous to the line ratios used in the VO87 diagram \citep{veil87}, except that the low spectral resolution  of the grism blends the \Ha\ and \NII$\lambda$6584 lines. We also define an ``OHNO'' sample of galaxies with $\mathrm{SNR}>1$ emission-line detections of \OIII$\lambda$5007, \Hb, \NeIII$\lambda$3870, and the blended \OII$\lambda$3726+3728 doublet. In both samples, the low grism resolution blends the \OIII$\lambda$4959+5007 doublet, and so we measure the \OIII$\lambda$5007 line as 75\% of the total (using the typical doublet ratio, \citealt{stor00}). We visually inspect the direct image, 1D, and 2D spectra of galaxies selected for these samples to ensure good emission-line detections and avoid spectra that are heavily contaminated by nearby objects, an example of one is shown in Figure \ref{fig:spec}.

These selection criteria result in 461 unVO87 galaxies in the redshift range $0.6<z<1.5$ and 91 OHNO galaxies in the redshift range $1.2<z<2.5$. The star formation rates (SFR) and stellar mass distributions of these samples are shown in Figure \ref{fig:sample}. 

The SNR of the emission lines and emission-line ratios of our samples is shown in Figure \ref{fig:SNR}. Although the minimum threshold for a line detection is only $\mathrm{SNR}>1$, 99\% (457/461) of galaxies in the unVO87 and 99\% (90/91) of galaxies in the OHNO sample have at least one emission line with $\mathrm{SNR}>3$. In particular, \OIII\ is usually detected with $\mathrm{SNR}>3$ in both samples: 82\% (378/461) in unVO87 and 92\% (84/91) in OHNO. In other words, almost all galaxies in our sample have at least one well-detected emission line and thus a secure spectroscopic redshift, such that marginal ($1<\mathrm{SNR}<3$) detections of weaker lines still provide valuable constraints on the emission-line ratios.

We use the luminosity of \Hb\ as a proxy for SFR, since it is detected in both galaxy samples. The \Hb\ SFR is found by following the \citet{kenn12} SFR relation for \Ha\ and $\Ha/\Hb=2.86$ (assuming Case B recombination, $T=10^4$~K, and $n_e=10^4$~cm$^{-3}$; \citealp{oste89}):

\begin{equation} \label{Eq:SFR}
  \log({\rm SFR})[M_\odot/\mathrm{yr}] = \log[L(\Hb)] - 40.82
\end{equation}
This SFR is not corrected for dust attenuation, and so is likely a lower limit on the true star formation in each galaxy.

We additionally identify \NeV-detected galaxies, selected as \NeV\ emission-line detections of SNR$>$2.
Spectra with \NeV\ detections were visually inspected in the same process as the galaxy samples for the unVO87 and OHNO diagrams. The \NeV\ ion requires very high ionization ($E=97$~eV) that is likely caused by AGN or energetic shocks rather than by typical star-forming \HII\ regions \citep{izot12,madd18}.

\subsection{Low Redshift Comparison Sample: SDSS}

We construct a low-redshift comparison sample of galaxies from
the Sloan Digital Sky Survey(SDSS) \citep{york00} Data Release 10 \citep{Ahn14}. The SDSS data set uses a 2.5-m telescope at Apache Point Observatory to cover 14,555 $deg^2$ in the sky. SDSS spectra have $R \sim 2000$ over $3800<\lambda<9200\AA$ \citep{smee13}.

Emission-line measurements and redshifts for the SDSS data set are computed by \cite{bolt12}, using a stellar template to correct the continuum for stellar absorption. Stellar masses are estimated by \cite{Mont16} from the broadband $ugriz$ SDSS photometry using a grid of templates made from the FSPS stellar population synthesis code \citep{conr09}. These templates assume a \cite{krou01} initial mass function (IMF) and fit for the dust attenuation following \cite{Char00} and \cite{calz00}.

We select the low-redshift comparison sample using the same $\mathrm{SNR}>1$ line detection thresholds as for the CLEAR samples. The SDSS spectra resolve most of the lines that are blended in the lower resolution grism spectra. We add the resolved \Ha\ and \NII\ line fluxes from SDSS together to mimic the blended version in the CLEAR data.

These selection criteria result in 245,242 unVO87 galaxies within the redshift range $0.01<z<0.1$ and 27,972 OHNO galaxies in the redshift range $0.025<z<0.1$ (with a slightly higher low-redshift cutoff to include \OII\ in the observed-frame spectrum).

\section{unVO87 and OHNO Diagrams to classify Galaxies}

We use the \OIII/\Hb, \SII/\HavNII, and \NeIII/\OII\ emission-line ratios to investigate the physical conditions of CLEAR galaxies. This section presents the general line-ratio properties of our CLEAR galaxy sample and defines lines to separate AGN from star-forming galaxies in both the unVO87 and OHNO diagrams.

\subsection{unVO87 diagram}

\begin{figure}[t]
\epsscale{1.2}
\plotone{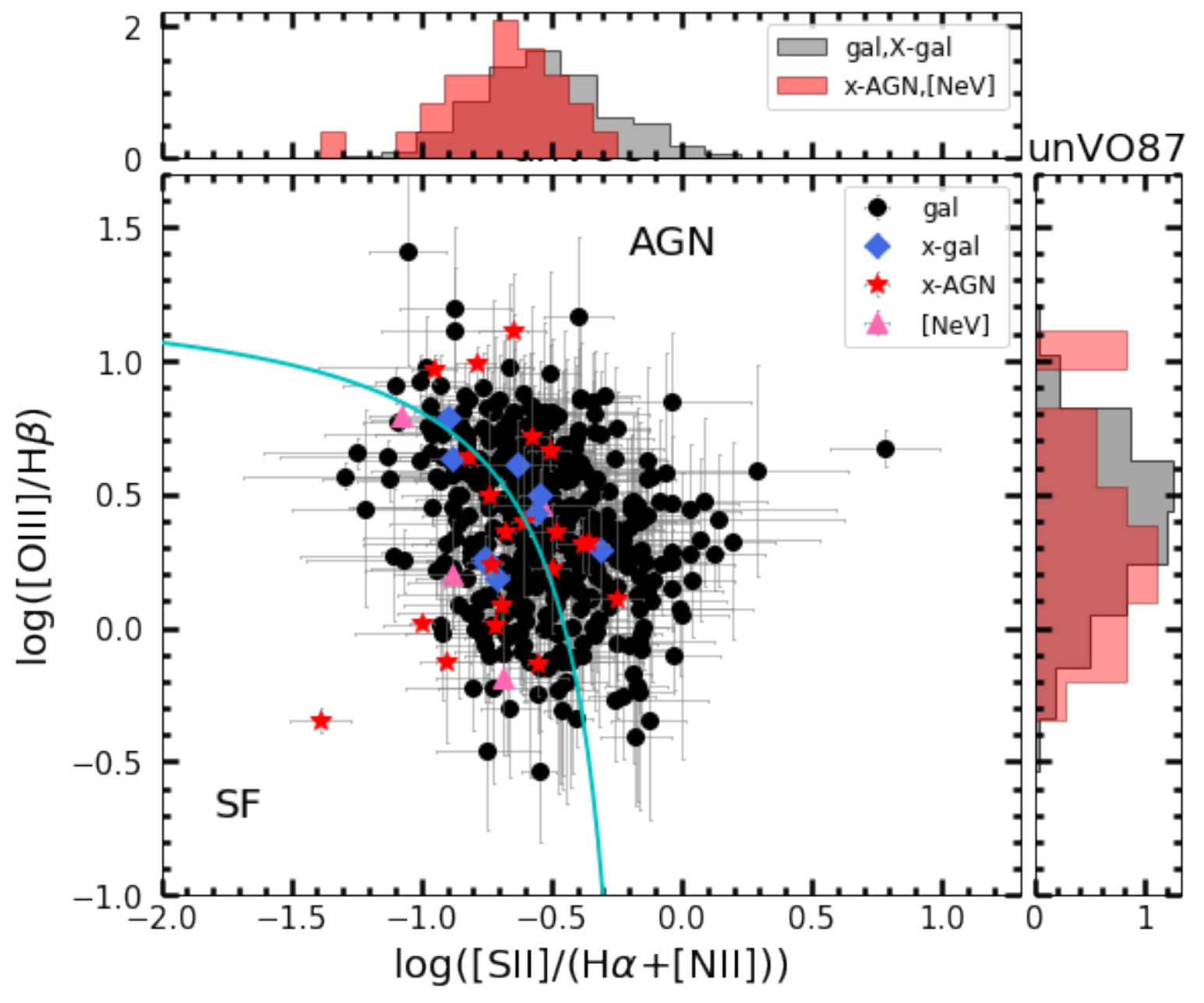}
\caption{The ``unVO87'' diagram of log(\OIII/\Hb) vs log(\SII/\HavNII) for the CLEAR samples. The cyan line shows the modified AGN/SF line given by Equation \ref{eq:unVOline}. Different color symbols indicate X-ray AGN, X-ray galaxies, and \NeV\ sources. The histograms show X-ray AGN and [NeV] in red and other galaxies in gray.
\label{fig:unVO87}}
\end{figure}

The emission-line ratios of our unVO87 sample of CLEAR galaxies are shown in Figure \ref{fig:unVO87}. Of these 461 galaxies, 20 are X-ray AGN, 8 are X-ray galaxies, 4 are $\NeV$ sources, and 1 is both an X-ray AGN and $\NeV$ detection.

\begin{figure}[t]
\epsscale{1.1}
\plotone{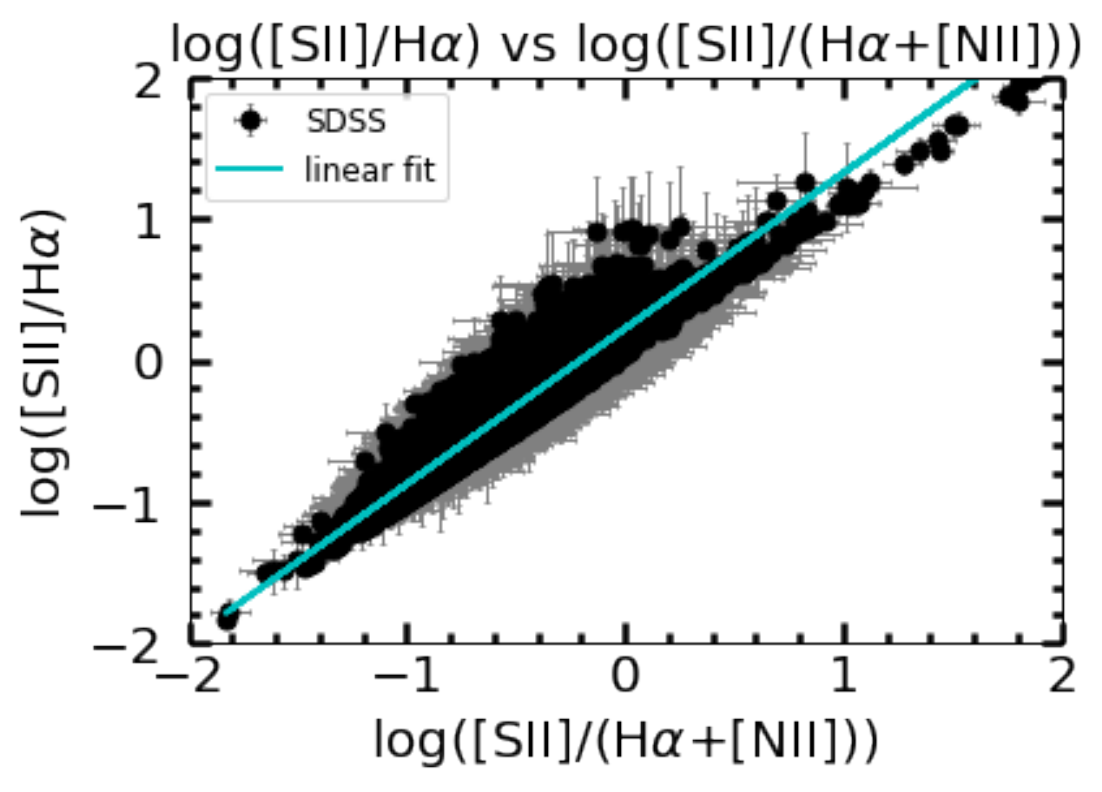}
\caption{The relationship between $\log(\SII/\Ha)$ and $\log(\SII/\HavNII)$ for our low-redshift SDSS galaxy sample. The cyan line indicates the best-fit linear regression. We use this best-fit line, $\log(\SII/\HavNII)=1.1\log(\SII/\Ha)+0.22$, to translate the VO87 AGN/SF dividing line into an analogous AGN/SF division in the unVO87 diagram. The new AGN/SF dividing line for the unVO87 diagram is given by Equation \ref{eq:unVOline}.
\label{fig:unVO87line}}
\end{figure}

The cyan line in Figure \ref{fig:unVO87} indicates the AGN/SF dividing line, separating galaxies with emission lines dominated by AGN above the line and galaxies with emission lines dominated by star formation activity below. We create this AGN/SF line starting from the similar line determined for the VO87 diagram in \citet{Trump15}, which was empirically designed to be parallel to the sequence of star-forming galaxies at $z<0.1$ observed by SDSS. We transform the VO87 line into our unVO87 diagram, with its unresolved \HavNII\ lines, by comparing the relationship between $\SII/\Ha$ and $\SII/\HavNII$ in SDSS galaxies. This comparison is shown in Figure \ref{fig:unVO87line}.

The cyan line shows the best-fit linear regression line to the line ratios, with the relationship $\log(\SII/\Ha)=1.1\log(\SII/\HavNII)+0.22$. We combine this relation with the VO87 line of \citet{Trump15} to find the modified unVO87 AGN/SF line in Figure \ref{fig:unVO87}, 

\vspace{0.1cm}
\begin{equation}\label{eq:unVOline}
\log\left(\frac{\OIII}{\Hb}\right)=\frac{0.48}{1.09*\log\left(\frac{\SII}{\HavNII}\right)+0.12}+1.3
\end{equation}

The $z \sim 1$ galaxies in our CLEAR sample have a broad distribution in the unVO87 diagram. Furthermore, the X-ray AGN and \NeV\ high-ionization galaxies occupy a very similar distribution to the larger sample, suggesting that the unVO87 AGN/SF line is not particularly effective at classifying $z \sim 1$ galaxies. We perform Anderson-Darling (A-D) tests to determine if the X-ray AGN and \NeV\ galaxies have the same parent distribution as the rest of the galaxy sample.
We use a null-probability of $p < 0.05$ as the threshold to distinguish that the two populations are not consistent with the same parent distribution. The A-D test for the $\log{(\SII/\HavNII)}$ emission line ratio finds a $p=0.0037$ for the population of X-ray AGN and \NeV\ galaxies compared to the population of other galaxies, showing the two populations are likely to come from different parent distributions. The A-D test for the $\log{(\OIII/\Hb)}$ emission line ratio finds a $p$-value greater than 0.25, suggesting the X-ray AGN and \NeV\ galaxies are consistent with sharing the same parent distribution as the rest of the galaxy sample.


\begin{figure}[t]
\epsscale{1.2}
\plotone{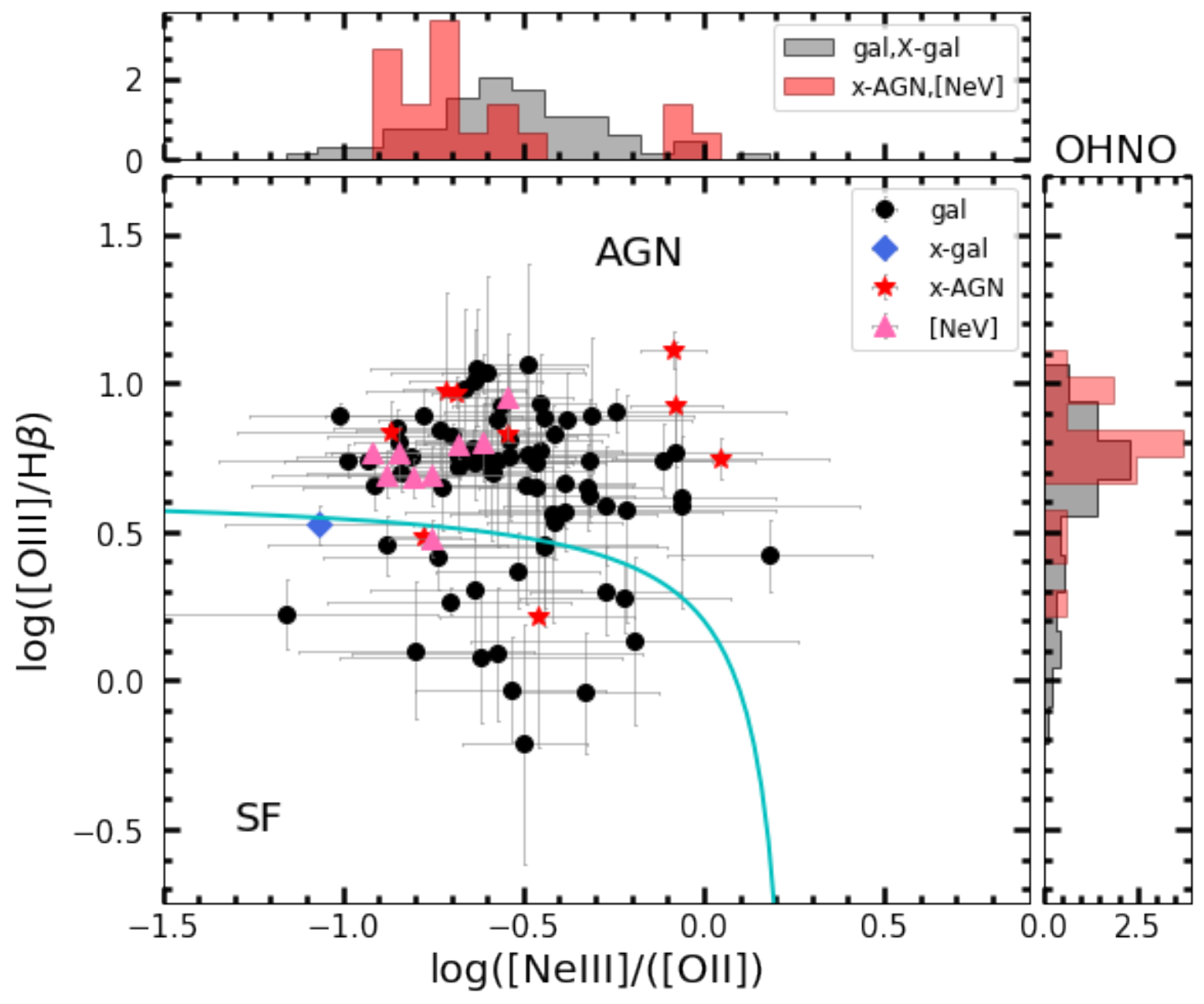}
\figcaption{The OHNO diagram showing log(\OIII/\Hb) vs log(\NeIII/\OII). The cyan line is an empirical AGN/SF dividing line defined in Equation \ref{Eq:OHNO line}. Different color symbols indicate X-ray AGN, X-ray galaxies, and \NeV\ sources. The histograms show the $\NeIII/\OII$ and $\OIII/\Hb$ distributions of the X-ray AGN and \NeV\ sources in red with the remaining galaxies shown in grey.X-ray AGN and high-ionization \NeV\ galaxies are more likely to sit above the cyan line than the rest of the galaxy population.
\label{fig:OHNO}}
\end{figure}

\begin{figure*}[t]
\epsscale{0.9}
\plotone{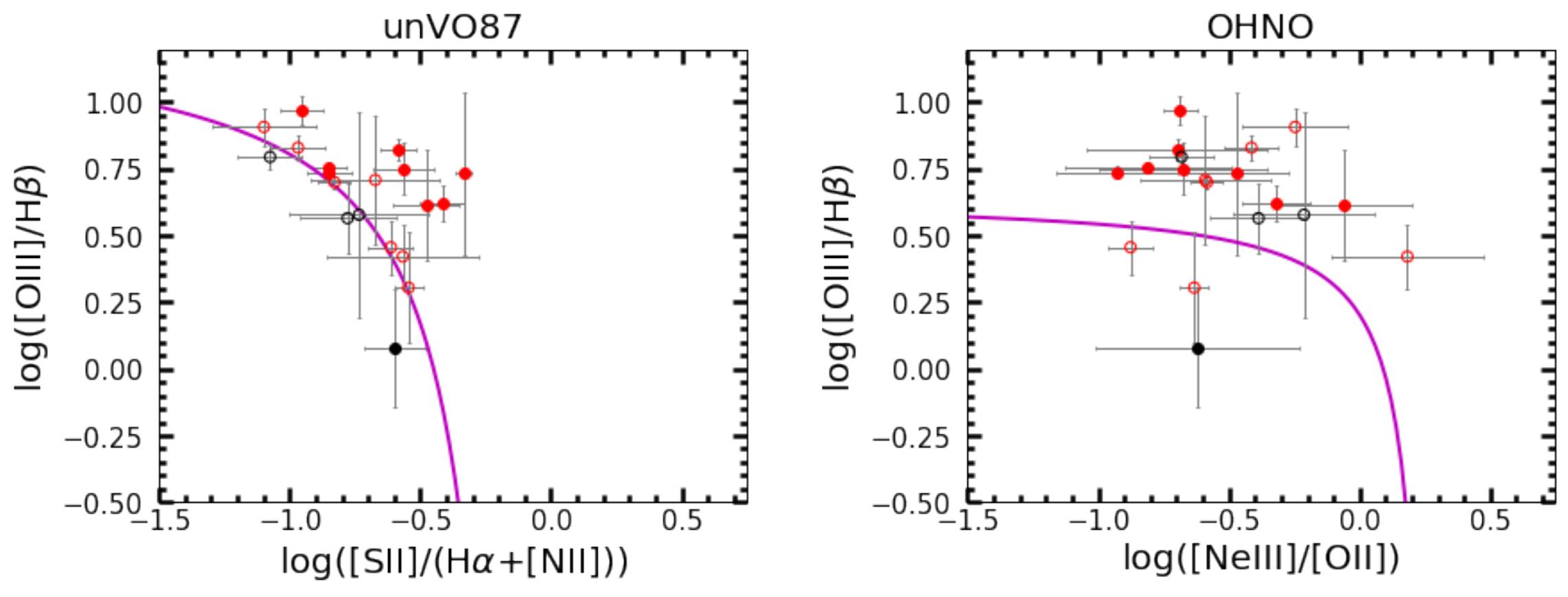}
\figcaption{The unVO87 and OHNO diagrams for the 19 galaxies at $1.2<z<1.5$ that are in both samples. The red and black colors indicate galaxies above and below the unVO87 AGN/SF line (Equation 2), respectively, with open symbols for galaxies with 1-$\sigma$ error bars that overlap with the line.The OHNO AGN/SF line is created to empirically produce similar separation between AGN and SF galaxies classified by the unVO87 diagram.
\label{fig:OHNOline}}
\end{figure*}

The lack of separation of X-ray AGN and high-ionization \NeV\ from the rest of the galaxy population in the unVO87 diagram contrasts with the VO87 diagram, which has long been shown to effectively separate AGN and inactive galaxies at $z \sim 0$ \citep[e.g.,][]{veil87,kauf03,kewl01,Kewl19}, and the unVO87 and VO87 line ratios are generally very similar (as shown in Figure \ref{fig:unVO87line}).

This might indicate redshift evolution of star-forming galaxies in the unVO87 and VO87 diagrams, such that there is significant overlap in the line ratios of AGN and high-ionization sources with non-AGN and moderate ionization galaxies at higher redshift. \citet{coil15} also found similar (resolved) VO87 line ratios for X-ray AGN and inactive galaxies in a sample of 56 galaxies at $z \sim 2.3$ from the MOSDEF survey. We further investigate the redshift evolution of galaxies in the unVO87 diagram in Section 4.

\subsection{OHNO diagram}

The ratio of the $\NeIII\lambda3870$ and $\OII\lambda3726+3729$ emission lines is an effective ionization diagnostic that is available for high-redshift galaxies \citep[e.g.,][]{trou11,Zeim15}. \NeIII\ has similar ionization potential to \OIII\ and is sufficiently close to the $\OII\lambda3727$ line that the ratio is largely insensitive to dust attenuation.

The line ratios of the 91 galaxies in our OHNO sample are shown in Figure \ref{fig:OHNO}. Of these galaxies, 4 are X-ray AGN, 1 is an X-ray galaxy, 8 are $\NeV$ sources, and 5 are both X-ray AGN and $\NeV$ sources. The cyan AGN/SF division line in Figure \ref{fig:OHNO} is created by comparing the 19 galaxies in our CLEAR dataset that are in both the unVO87 and OHNO samples. These galaxies are plotted in the unVO87 and OHNO diagrams in Figure \ref{fig:OHNOline} and are color-coded by their position above or below the unVO87 AGN/SF division line defined by Equation \ref{eq:unVOline}. We define an AGN/SF line in the OHNO diagram to empirically match the unVO87 classifications, given by the equation: 

\begin{equation} \label{Eq:OHNO line}
  \log\left(\frac{\OIII}{\Hb}\right)=\frac{0.35}{2.8\log\left(\frac{\NeIII}{\OII}\right)-0.8}+0.64
\end{equation}

The OHNO AGN/SF line is chosen such that the 8 galaxies that are above the unVO87 AGN/SF line also lie above the OHNO line, and the 1 unVO87 SF galaxy is similarly classified as a OHNO SF galaxy. Of the 11 unVO87 galaxies that have 1-$\sigma$ uncertainties that overlap with the AGN/SF dividing line, 9 are above the OHNO AGN/SF line and 2 are below it.

In Figure \ref{fig:OHNO}, the \NeV\ sources tend to prefer higher $\OIII/\Hb$ and lower $\NeIII/\OII$ line ratios, whereas the rest of the sample spans a larger range of emission-line ratios. X-ray AGN are similarly well-separated by the dividing line, with 6 X-ray AGN above the line and only 2 falling (marginally, within 1$\sigma$) below it. This suggests that the OHNO diagram can be effective for distinguishing high-ionization galaxies and AGN. An A-D test finds a null probability $p=0.046$ that the $\log{(\NeIII/\OII)}$ emission-line ratios of X-ray AGN and \NeV\ galaxies are from the same parent distribution as the other galaxy line ratios, suggesting that the populations may be well-separated by $\log{(\NeIII/\OII)}$ at $z>1$. The A-D test for the $\log{(\OIII/\Hb)}$ emission line is $p=0.11$ showing again that X-ray AGN and \NeV\ galaxies are not inconsistent with the same parent distribution as the rest of the galaxies.
We further investigate the line-ratio evolution of galaxies with redshift in Section 4.

\section{Redshift Evolution of Galaxy Emission-Line Properties}

We use the $0.6<z<2.5$ CLEAR and $z<0.1$ SDSS galaxy samples to compare galaxy emission-line properties at high and low redshift.
We first split the CLEAR unVO87 and OHNO samples into low and high redshift bins using the average redshift of each sample, z=1.08 and z=1.69 respectively. This results in four samples of CLEAR emission-line galaxies: unVO87 samples at $z \sim 0.9$ and $z \sim 1.3$ and OHNO samples at $z \sim 1.4$ and $z \sim 2.0$.

We construct comparison samples of SDSS galaxies following the approach of \cite{june14}. We first define a luminosity detection threshold for CLEAR galaxies in each bin. Because the CLEAR samples require ${\rm SNR}>1$ emission-line detections, the typical detection threshold is given by the average line flux uncertainty in each bin.

\begin{figure*}[t]
\epsscale{1.2}
\plotone{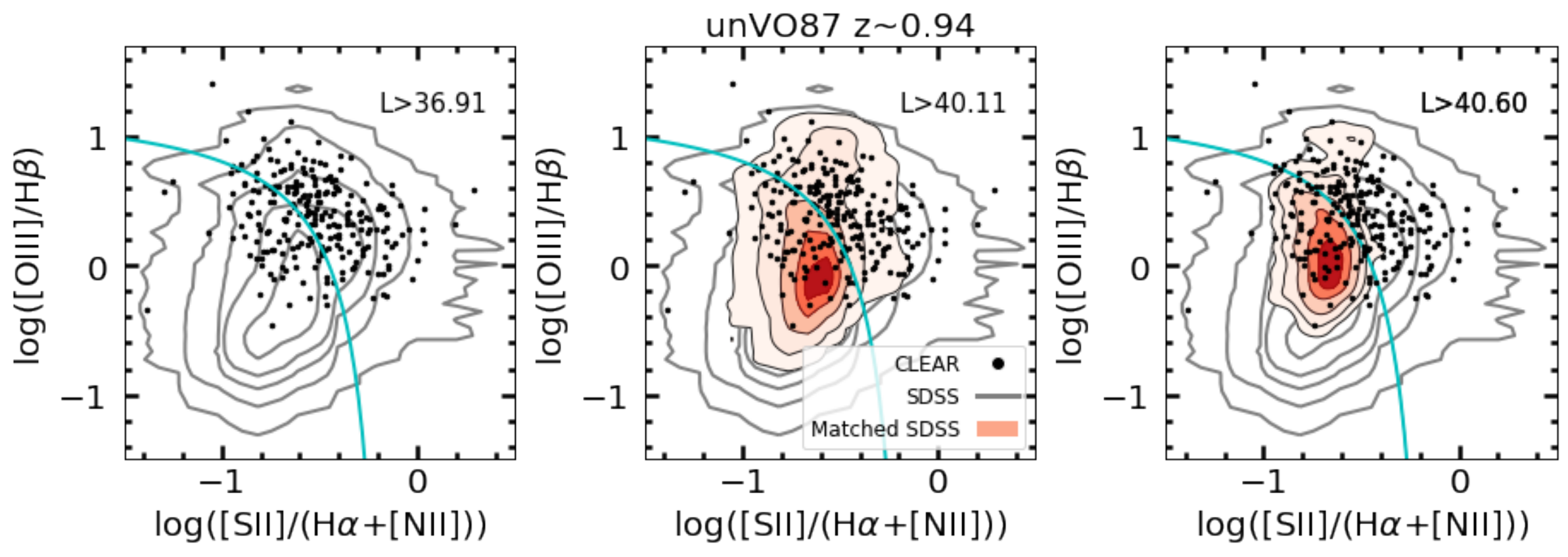}
\caption{The unVO87 line ratios for CLEAR galaxies in the low-redshift half of the unVO87 sample (black points, mean redshift $z=0.94$) compared to $z \sim 0$ SDSS galaxies (red and gray contours). The gray contours represent all SDSS galaxies with emission-line luminosities of $L>36.91$. The red contours in the center panel show the ``evolution-matched'' SDSS sample with luminosities above $L>40.11$, calculated for redshift evolution of emission-line luminosities following Equation 4. The red contours in the right panel show the ``luminosity-matched'' SDSS galaxies above the same $L>40.60$ emission-line luminosity detection threshold as the CLEAR galaxies at $z \sim 0.94$. The CLEAR galaxies have $\sim$0.5~dex higher average $\OIII/\Hb$ and $\sim$0.1~dex higher \SII/\HavNII\ than the low-redshift SDSS galaxies. 
\label{Fig:unvo87levo}}
\end{figure*}

\begin{figure*}[t]
\epsscale{1.2}
\plotone{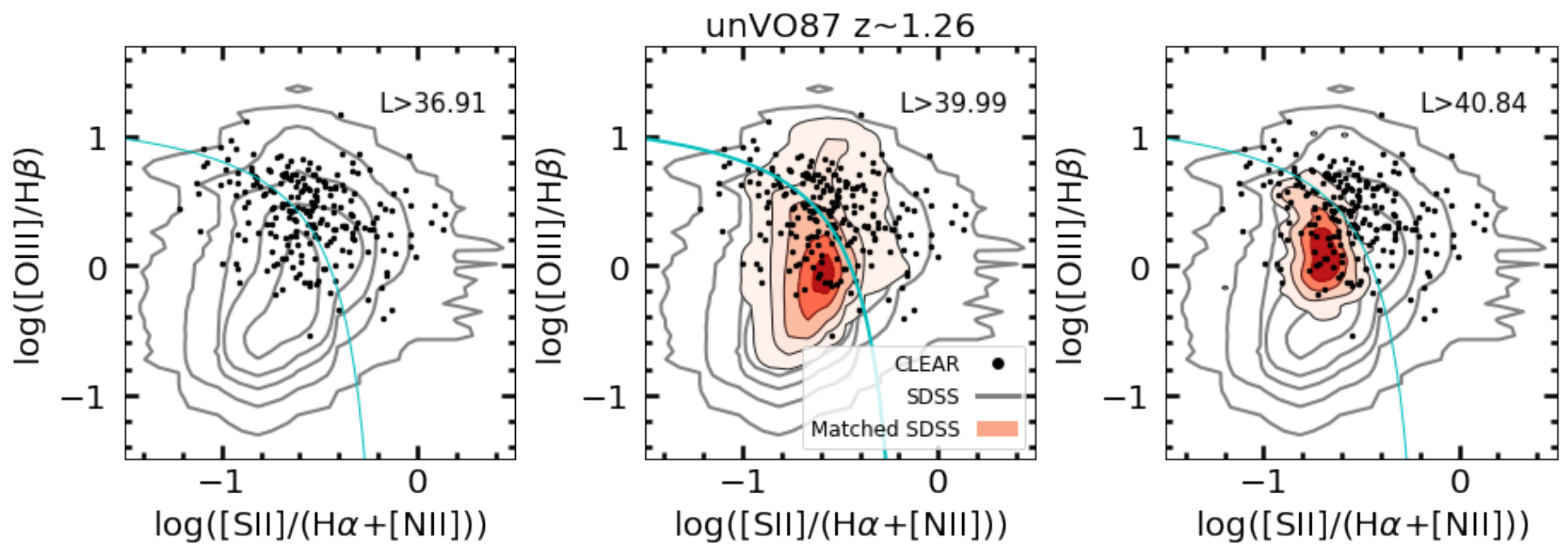}
\caption{The unVO87 line ratios for CLEAR galaxies in the high-redshift half of the unVO87 sample (black points, mean redshift $z=1.26$) compared to $z \sim 0$ SDSS galaxies (red and gray contours). The gray contours represent all SDSS galaxies with emission-line luminosities above $L>36.91$. Red contours indicate ``evolution-matched'' SDSS galaxies with $L>39.99$ for all four emission lines (center panel), and `luminosity-matched'' SDSS galaxies with $L>40.84$ for all four emission lines (right panel). As in Figure \ref{Fig:unvo87levo}, the high-redshift CLEAR galaxies tend to have higher $\OIII/\Hb$ and \SII/\HavNII\ and more frequently lie above the AGN/SF line than the low-redshift SDSS galaxies.
\label{Fig:unvo87hevo}}
\end{figure*}

For each redshift bin of CLEAR galaxies, we define a ``luminosity-matched'' low-redshift sample of SDSS galaxies that have all four lines (unVO87 or OHNO) with luminosities greater than the luminosity detection threshold of the CLEAR galaxies. Galaxies in the unVO87 diagram are usually limited by the \SII\ or \Hb\ lines and galaxies in the OHNO diagram are usually limited the \NeIII\ or \Hb\ lines. We also define an ``evolution-matched'' sample of low-redshift galaxies, assuming evolution in emission-line luminosities that matches the evolution of the star formation mass sequence \citep{whit12}:

\begin{equation} 
\begin{split}
  \log{L(0)} = \log{L(z)} + 0.7 \mathcal{M}(0)  - (0.7-0.13z) \mathcal{M}(z) & - 1.14z + 0.19z^2 \\ \mathcal{M}(z)=[\log{\Ms(z)}-10.5]
\end{split}
\label{eq:zcor}
\end{equation}


We use the average redshift and stellar mass in each bin of CLEAR galaxies to evaluate the line luminosity evolution in this equation, solving for the corresponding $z=0$ luminosity to define the evolution-matched SDSS samples. The unVO87 SDSS sample has an average stellar mass of $\log{\Ms(0)}=10.27$ and the OHNO SDSS sample has an average stellar mass of $\log{\Ms(0)}=10.02$.

\begin{figure}[t]
\epsscale{1.2}
\plotone{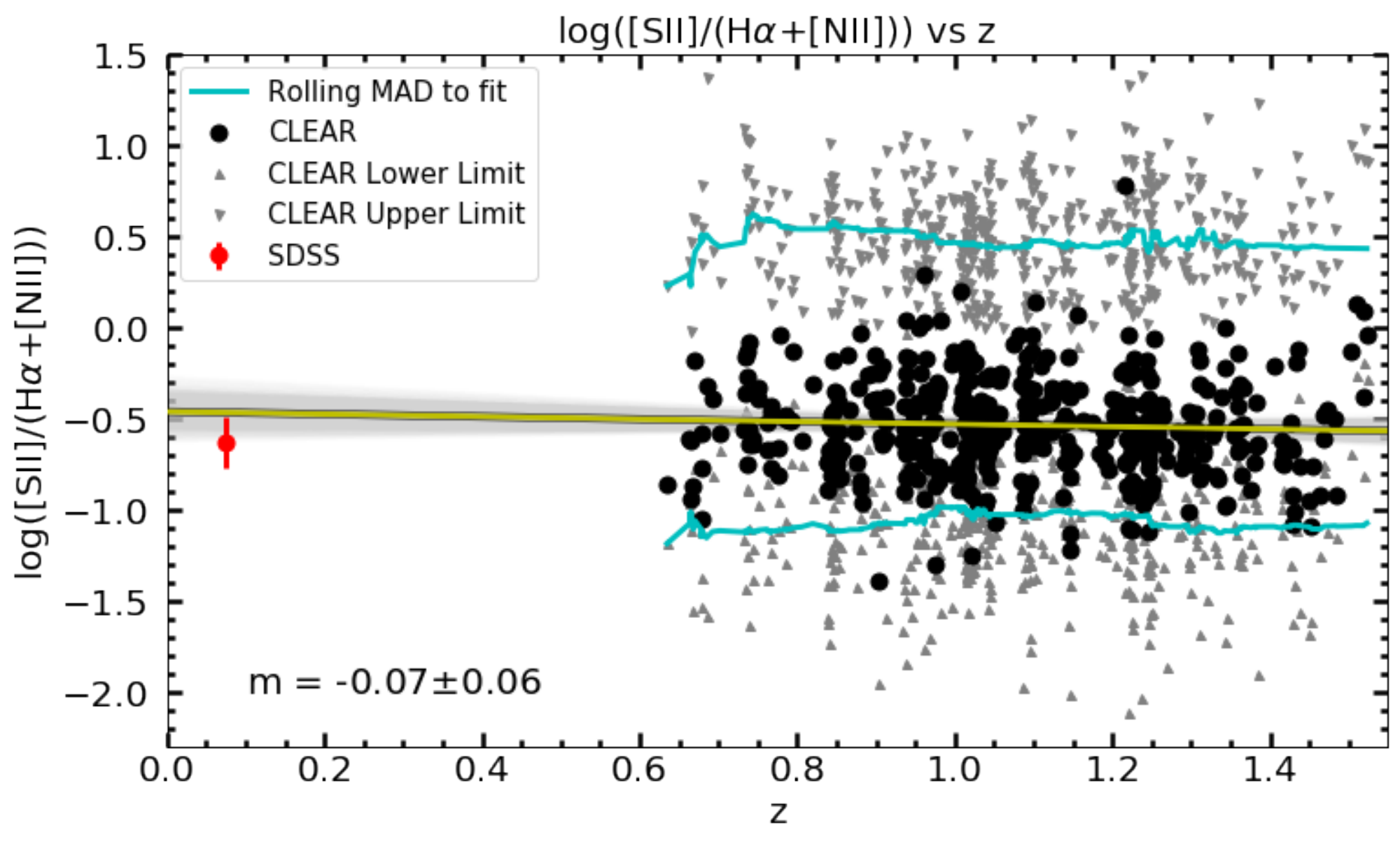}
\figcaption{The evolution of the \SII/\HavNII\ line ratio with redshift. Black points show CLEAR galaxies, with upper and lower detection limits indicated by gray triangles. The yellow line is the best-fit linear regression to the CLEAR data points. The cyan line shows the rolling MAD of the CLEAR detection limits to the best-fit line. The red point indicates the mean line ratio of the ``evolution-matched'' SDSS sample, calculated using Equation \ref{eq:zcor} and the median redshift of the CLEAR sample, with red error bars indicating the standard deviation of the SDSS line ratios in the sample.} The blended \SII/\HavNII\ line ratio does not significantly evolve with redshift (-0.07$\pm$0.06).
\label{Fig:S2vsz}
\end{figure}

Figure \ref{Fig:unvo87levo} and Figure \ref{Fig:unvo87hevo} show the line ratios of the low-redshift ($z \sim 0.9$) and high-redshift ($z \sim 1.3$) CLEAR unVO87 samples, respectively, compared to $z \sim 0$ SDSS galaxies. In each figure, black points indicate the CLEAR galaxies and red contours indicate the matched SDSS samples. The left panels use no matching and show the full SDSS sample as gray contours, 
the center panels show the ``evolution-matched'' SDSS galaxy samples using the luminosity evolution given by Equation \ref{eq:zcor} evaluated at the average redshift of the CLEAR galaxies in each bin, and the right panels show the ``luminosity-matched'' SDSS samples that have the same line luminosity limit (with no evolution) as the CLEAR galaxies.

Figures \ref{Fig:unvo87levo} and \ref{Fig:unvo87hevo} show that the higher redshift CLEAR galaxies generally have higher $\OIII/\Hb$ than the $z \sim 0$ SDSS galaxies. Most of the CLEAR galaxies lie above the unVO87 AGN/SF dividing line, while most galaxies in the matched SDSS samples lie below the line. Compared to the SDSS sample, the CLEAR galaxies generally have $\SII/\HavNII$ that is $\sim$0.1~dex higher and $\OIII/\Hb$ that is $\sim$0.5~dex higher.

\begin{figure}[t]
\epsscale{1.2}
\plotone{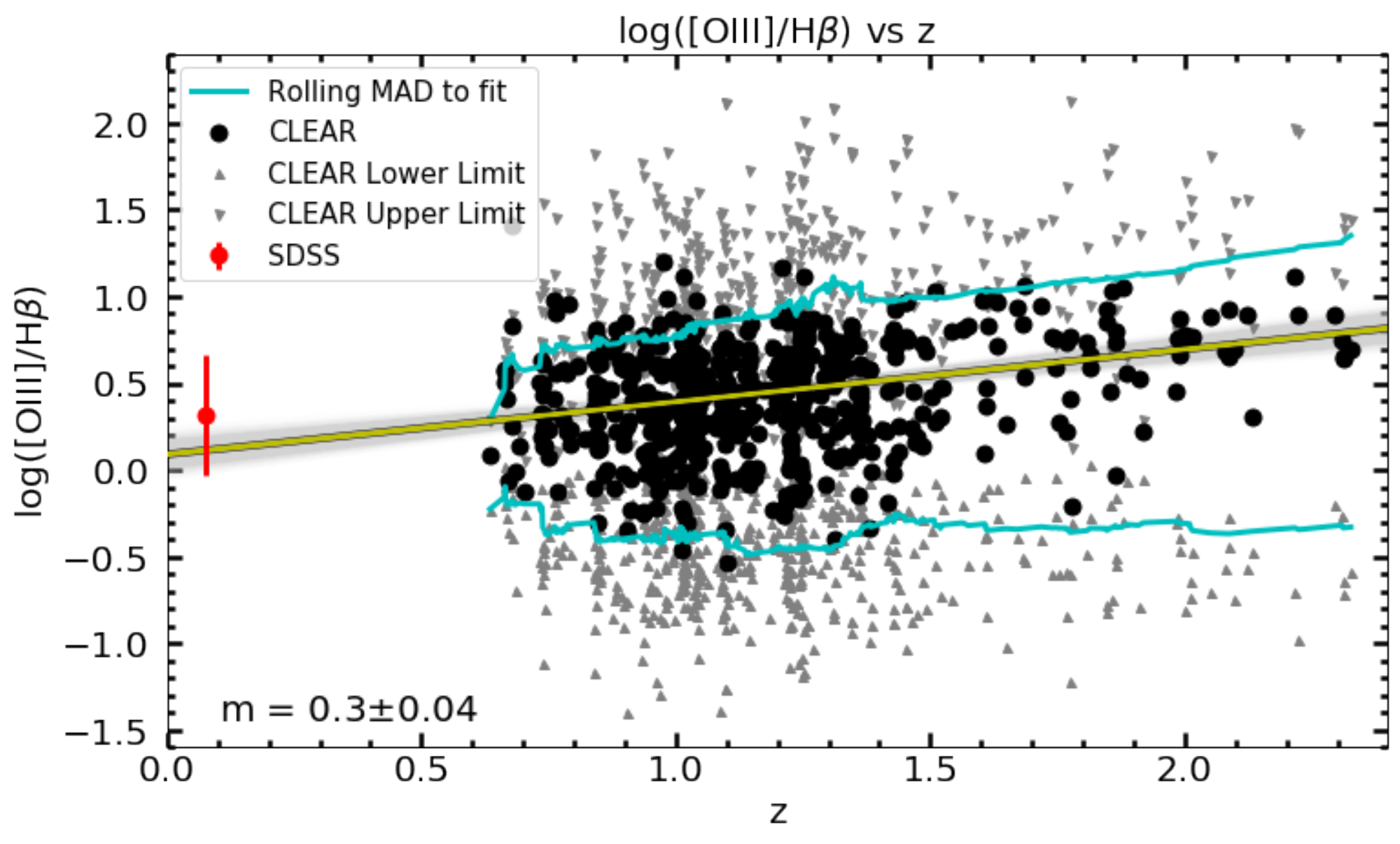}
\figcaption{The evolution of the \OIII/\Hb\ line ratio with redshift. CLEAR galaxies are indicated by black points, accompanied by gray triangles to indicate the upper and lower detection limits for each galaxy. The yellow line is the best-fit linear regression to the CLEAR data points. The cyan line shows the rolling MAD of the CLEAR detection limits to the best fit line. The red point indicates the mean line ratio of the ``evolution-matched'' SDSS sample, calculated using Equation \ref{eq:zcor} and the median redshift of the CLEAR sample, with red error bars indicating the standard deviation of the SDSS line ratios in the sample. The \OIII/\Hb\ emission-line ratio increases with redshift, with a best-fit linear regression slope (0.3$\pm$0.04) that is significantly different from zero.
\label{Fig:O3vsz}}
\end{figure}

We further examine the redshift evolution of the unVO87 line ratios in Figures \ref{Fig:S2vsz} and \ref{Fig:O3vsz}. The \SII/\HavNII\ line ratio (shown in Figure \ref{Fig:S2vsz}) does not significantly evolve with redshift, with a best-fit linear regression slope that is consistent with zero among the CLEAR data, and similar line ratios for both SDSS and CLEAR galaxies. On the other hand, the \OIII/\Hb\ line ratio has a significant correlation (slope of 0.30 $\pm$ 0.04) with redshift among the CLEAR galaxies, along with a significant difference in the average \OIII/\Hb\ line ratio between CLEAR and SDSS $z \sim 0$ galaxies. The increase of \OIII/\Hb\ at higher redshift does not appear to be solely driven by selection effects (indicated by the gray triangles and cyan lines in Figures \ref{Fig:S2vsz} and \ref{Fig:O3vsz}). There may be some high-\OIII/\Hb\ galaxies at $z \sim 1$ that are missed due to the SNR$>$1 requirement for \Hb, but there also seems to be a genuine lack of low-\OIII/\Hb\ galaxies at $z>1.5$ that would be detectable above the \OIII\ detection limits. We thus conclude that the unVO87 line ratio differences between low-redshift and high-redshift galaxies are principally driven by increasing \OIII/\Hb\ ratios in higher redshift galaxies. At fixed stellar mass, high-redshift galaxies tend to have higher SFRs, which is correlated with higher \OIII/\Hb\ ratios and higher ionization (Papovich et al., in prep). We further investigate the dependence of \OIII/\Hb\ ratios on galaxy properties in Section 5.1. 

\begin{figure*}[t]
\epsscale{1}
\plotone{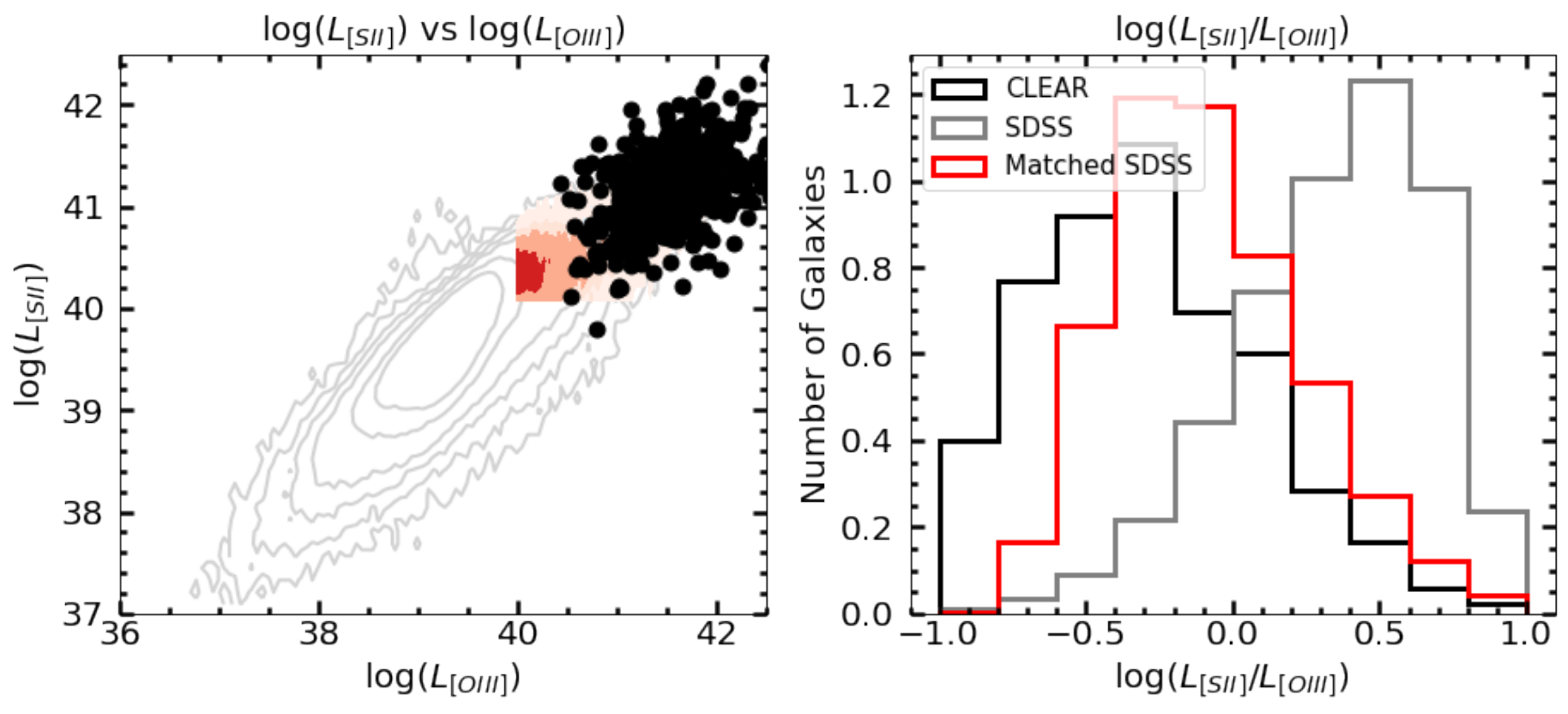}
\figcaption{Comparison of the \SII\ and \OIII\ luminosities for CLEAR galaxies (black points and histogram) and SDSS galaxies (gray contours and histogram). The red contours and red histogram show the ``evolution-matched'' SDSS sample calculated to match the mean redshift of the CLEAR sample using Equation 4. There is significant overlap of the line luminosities in the CLEAR and SDSS galaxies, although the right panel shows that high-redshift CLEAR galaxies tend to have slightly weaker \SII\ emission at fixed \OIII\ luminosity.
\label{Fig:S2vsO3}}
\end{figure*}

\begin{figure*}[t]
\epsscale{1.2}
\plotone{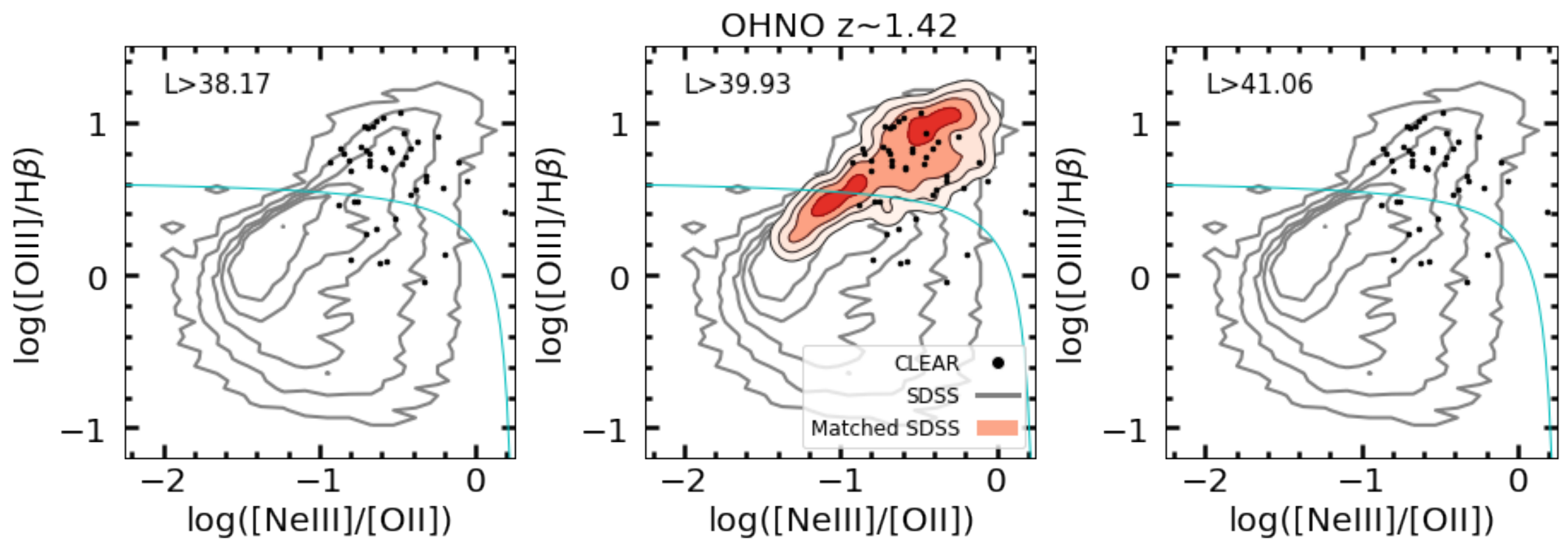}
\figcaption{The OHNO line ratios for CLEAR galaxies in the low-redshift half of the OHNO sample (black points, mean redshift $z=1.42$) compared to $z \sim 0$ SDSS galaxies (red and gray contours). The gray contours represent all SDSS galaxies with luminosities above $L>38.17$. The red contours in the center panel show the ``evolution-matched'' SDSS sample with emission-line luminosities above $L>39.93$, calculated for redshift evolution of emission-line luminosities following Equation 4. The right panel includes no matched SDSS galaxies because none of them have all four emission lines above the CLEAR luminosity detection threshold of $L>41.06$.
The $z \sim 1.42$ CLEAR galaxies tend to have higher \NeIII/\OII\ at fixed \OIII/\Hb\ than the low-redshift SDSS galaxies.
\label{Fig:OHNOlevo}}
\end{figure*}

\begin{figure*}[t]
\epsscale{1.2}
\plotone{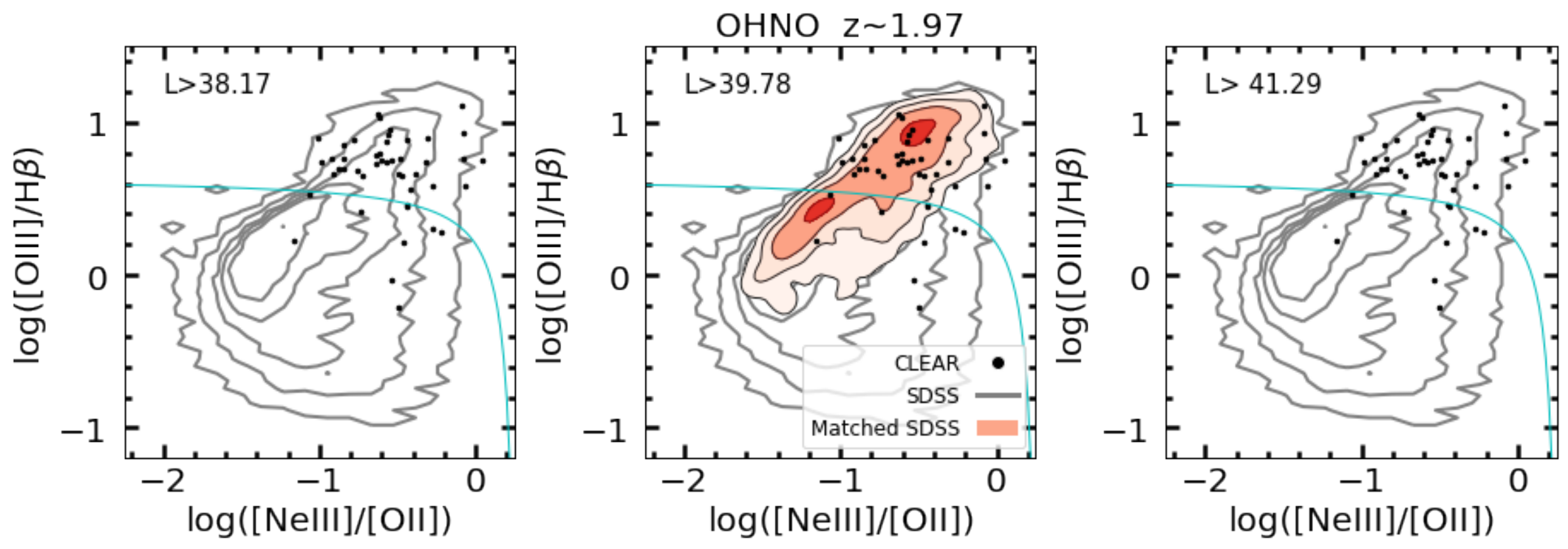}
\figcaption{The OHNO line ratios for CLEAR galaxies in the high-redshift half of the OHNO sample (black points, mean redshift $z=1.97$) compared to $z \sim 0$ SDSS galaxies (red and gray contours). The gray contours represent all SDSS galaxies with emission-line luminosities above $L>38.17$. The red contours in the center panel show the ``evolution-matched'' SDSS sample with emission-line luminosities above $L>39.78$, calculated for redshift evolution of emission-line luminosities following Equation 4. The right panel includes no matched SDSS galaxies because none of them have all four emission lines above the CLEAR luminosity detection threshold of $L>41.06$.
As in Figure \ref{Fig:OHNOlevo}, the CLEAR galaxies tend to have higher \NeIII/\OII\ at fixed \OIII/\Hb\ than the low-redshift SDSS galaxies.
\label{Fig:OHNOhevo}}
\end{figure*}

\begin{figure}[t]
\epsscale{1.2}
\plotone{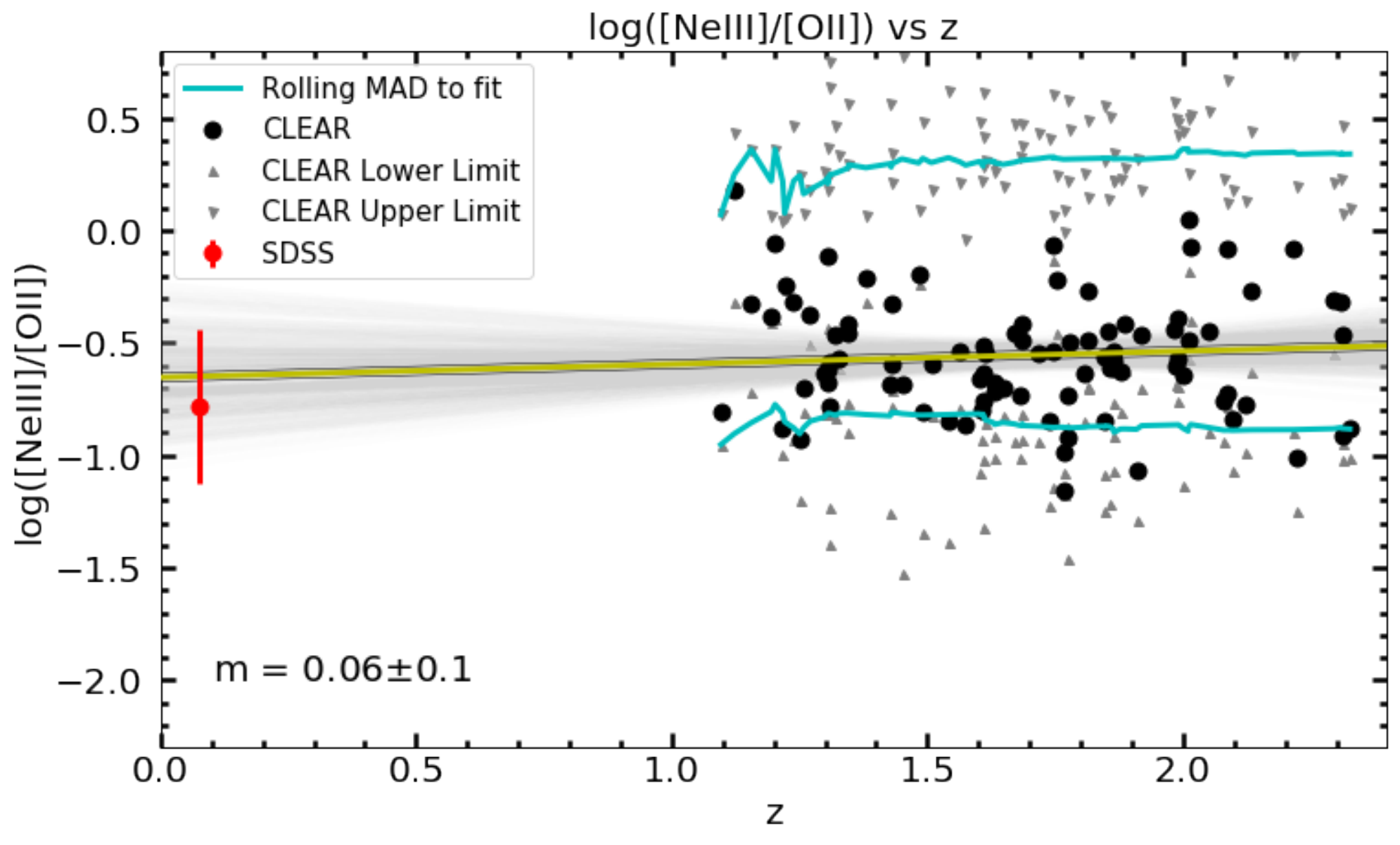}
\figcaption{The $\log{(\NeIII/\OII)}$ emission line ratio versus redshift. The CLEAR data are shown as black points with the grey triangles on either side indicating the lower and upper limits for galaxies to be detected above the SNR$>$1 threshold of our sample. The yellow line is the best-fit linear regression of the CLEAR data points. The cyan line shows the rolling MAD of the CLEAR detection limits to the best fit line. The red point indicates the mean line ratio of the ``evolution-matched'' SDSS sample, calculated using Equation \ref{eq:zcor} and the median redshift of the CLEAR sample, with red error bars indicating the standard deviation of the SDSS line ratios in the sample.
\label{Fig:Ne3vsz}}
\end{figure}

\begin{figure*}[t]
\epsscale{1}
\plotone{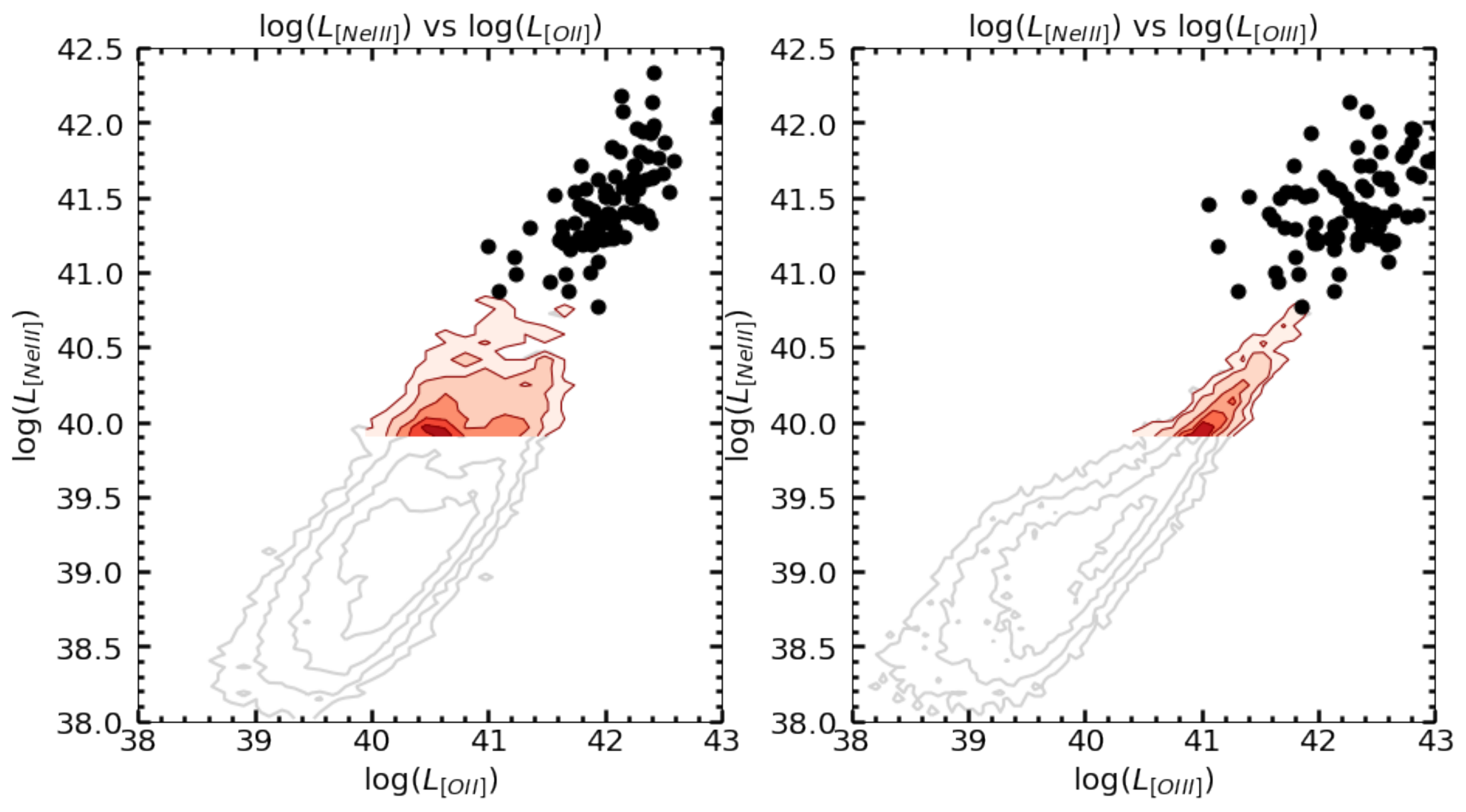}
\figcaption{Comparison of the \NeIII\ luminosity with the \OII\ and \OIII\ luminosities, with the black data points representing CLEAR data and the red contours representing SDSS galaxies in the ``evolution-matched'' sample with emission line luminosities above the threshold calculated from Equation 4. CLEAR and SDSS galaxies tend to have similar \NeIII/\OII\ line \textit{ratios}, 
as also seen in Figure \ref{Fig:Ne3vsz}. However both emission lines have significantly higher \textit{luminosity} at higher redshift
In the right panel, the high-redshift CLEAR galaxies also tend to have higher \NeIII\ luminosities at fixed \OIII\ luminosity compared to low-redshift SDSS galaxies.
\label{Fig:Ne3vsO3}}
\end{figure*}

\citet{Shap19} and \citet{Kash19} find that $1.3 \leq z \leq 2.7$ galaxies tend to have both lower \SII/\Ha\ and lower $\NII/\Ha$ ratios than $z \sim 0$ SDSS galaxies of the same \OIII/\Hb, suggesting that our blended \SII/\HavNII\ ratio might have redshift evolution in both the numerator and denominator that has a net result of a non-evolving \SII/\HavNII\ over $0<z<1.5$. We attempt to isolate the \SII\ line evolution by directly comparing the $\SII$ and $\OIII$ luminosities of CLEAR and SDSS galaxies in Figure \ref{Fig:S2vsO3}. Although the low-redshift SDSS and high-redshift CLEAR galaxies occupy similar parameter space in \SII\ and \OIII\ luminosities, the CLEAR galaxies frequently have lower \SII/\OIII\ luminosity ratios. The emission-line luminosities in Figure \ref{Fig:S2vsO3} are not corrected for dust attenuation, and correcting for dust would tend to increase the \OIII\ luminosity and decrease the \SII/\OIII\ ratio. Since dust attenuation is typically higher for $z \sim 1$ galaxies than $z \sim 0$ \citep[e.g.,][]{Burg13}, this would likely increase the difference between the low-redshift SDSS and high-redshift CLEAR galaxies. We therefore conclude that the high-redshift CLEAR galaxies tend to have weaker \SII\ emission at fixed \OIII\ luminosity, in agreement with the \citet{Shap19} and \citet{Kash19}.

Figures \ref{Fig:OHNOlevo}-\ref{Fig:OHNOhevo} display the comparisons between the OHNO line ratios of $z \sim 0$ SDSS galaxies with the CLEAR galaxies at $z \sim 1.4$ and $z \sim 2.0$, respectively. As in Figures \ref{Fig:unvo87levo} and \ref{Fig:unvo87hevo}, the three panels show the entire SDSS sample, the ``evolution-matched'' SDSS sample with all four emission lines exceeding the luminosity calculated from Equation \ref{eq:zcor}, and the ``luminosity-matched'' SDSS sample with all four lines exceeding the same luminosity limit as the CLEAR galaxies. There are no galaxies in the ``luminosity-matched'' panels of the two figures because no SDSS galaxies have $\NeIII$ emission lines as luminous as the CLEAR galaxies. Compared to the ``evolution-matched'' SDSS galaxies, high-redshift CLEAR galaxies tend to prefer higher $\NeIII/\OII$ (by $\sim$0.2~dex) at fixed $\OIII/\Hb$. This matches the result in \cite{Zeim15}.
Low-redshift SDSS galaxies do not span the same range of high $\NeIII/\OII$ ratios observed for high-redshift CLEAR galaxies, especially for galaxies with lower $\OIII/\Hb$. Interestingly, the evolution-matched SDSS galaxies are most common in two peaks above and below the OHNO AGN/SF dividing line, while high-redshift CLEAR galaxies instead most commonly have OHNO line ratios between the two low-redshift peaks.

Figure \ref{Fig:Ne3vsz} shows the \NeIII/\OII\ line ratio as a function of redshift for CLEAR galaxies (black points) and compared to the average line ratio of SDSS galaxies (red point). As in Figures \ref{Fig:S2vsz} and \ref{Fig:O3vsz}, the gray triangles show the detection limits for individual galaxies and the cyan line is a running median of these detection limits. The observed \NeIII/\OII\ line ratios of the CLEAR sample are likely to be significantly affected by the detection limit, since many of our observations lie just above the lower detection limit for the ratio and there might exist many $z>1$ galaxies with lower \NeIII/\OII\ that would lie below our SNR$>$1 detection threshold for \NeIII. Within the detection limits for our sample, there is no significant redshift evolution for the \NeIII/\OII\  line ratio.

Figure \ref{Fig:Ne3vsO3} directly compares the luminosity of the $\NeIII$ line with the $\OII$ and $\OIII$ lines for SDSS and CLEAR galaxies. As in the center panels of Figures \ref{Fig:unvo87levo}-\ref{Fig:OHNOhevo}, gray contours show the entire SDSS sample, red contours show the ``evolution-matched'' SDSS galaxies, and black points are CLEAR galaxies. Although CLEAR and SDSS galaxies have similar \NeIII/\OII\ line \textit{ratios} (as similarly seen in Figures \ref{Fig:OHNOlevo}-\ref{Fig:Ne3vsz}), the CLEAR galaxies are significantly more luminous in both lines, with no $z\sim 0$ SDSS galaxies that have \NeIII\ as luminous as the high-redshift CLEAR galaxies. There are even larger differences between SDSS and CLEAR galaxies in the comparison of their \NeIII\ and \OIII\ line luminosities. At fixed \OIII\ luminosity, high-redshift CLEAR galaxies tend to have more luminous \NeIII\ emission than (evolution-matched) low-redshift SDSS galaxies. The line luminosities in Figure \ref{Fig:Ne3vsO3} are not corrected for dust attenuation, but since attenuation is generally higher in high-redshift galaxies \citep[e.g.,][]{Burg13}, correcting for attenuation would likely increase the unusually high \NeIII/\OIII\ line ratios in high-redshift galaxies. \citet{Zeim15} similarly found anomalously luminous \NeIII\ emission in high-redshift galaxies and concluded that it is likely due to some combination of higher AGN content, different stellar populations (particularly Wolf-Rayet stars), and/or higher density \HII\ regions in high-redshift galaxies.

\begin{deluxetable*}{l|r|r|r|r}[t]
\tablecaption{Galaxies in each sample and redshift bin} \label{tab:mlinmix}
\tablenum{2}
\tablecolumns{5}
\tablewidth{0pt}
\tablehead{\colhead{} &\colhead{CLEAR}&\colhead{All SDSS} &\colhead{SDSS Evolution Matched}& \colhead{SDSS Luminosity Matched}}
\startdata
unVO87 $z \sim 0.94$ &  240 &       245242 &     8847 &      977 \\
unVO87 $z \sim 1.26$ & 222 &       245242 &     13868 &      448 \\
OHNO $z \sim1.42$ &  45 &       27972 &    190 &      0\\
OHNO $z \sim 1.97$  & 46 &       27972 & 934 &     0\\
\enddata
\end{deluxetable*}

\section{Interpretation of Emission Line Properties}\label{Properties}
 
\subsection{SFR and Stellar Mass}

We now investigate how the emission-line ratios correlate to galaxy stellar mass and SFR. 

\begin{figure*}[t]
\epsscale{1.1}
\plotone{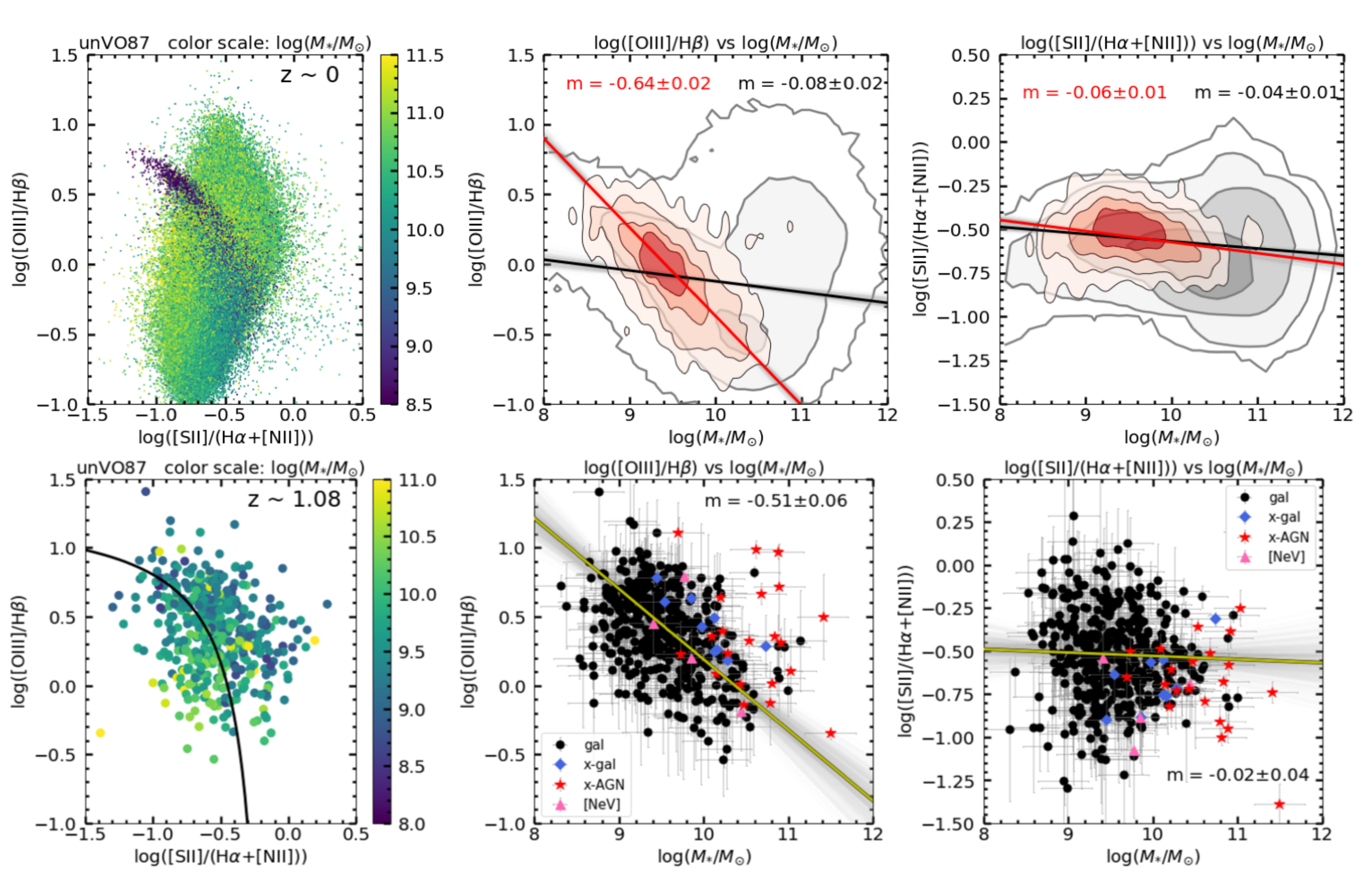}
\figcaption{The relationship between the unVO87 emission-line ratios and galaxy stellar mass. The top three panels depict the low-redshift SDSS unVO87 data and the bottom three panels depict the high-redshift CLEAR unVO87 data. The left panels display the unVO87 diagrams with galaxies color-coded by stellar mass, the middle panels display the relation of $\log{(\OIII/\Hb)}$ with stellar mass, and the right panels show the relation between $\log{(\SII/\HavNII)}$ and stellar mass. The middle and right panels include a best-fit linear regression line for the non-AGN (excluding the X-ray AGN and \NeV\ galaxies). The top middle and right panels include two fits and contours, grey/black to the entire SDSS sample and red for a subsample of SDSS galaxies that is matched to the same stellar mass distribution as the CLEAR galaxies. The high-redshift CLEAR $\log{(\OIII/\Hb)}$ ratios are significantly anti-correlated with stellar mass (bottom middle panel), and this same anti-correlation is present in the mass-matched SDSS sample (red contours top middle panel). Comparing the red SDSS contours to the CLEAR sample we also observe the $\log{(\OIII/\Hb)}$ is about 0.5 dex higher at high-redshift. The $\log{(\SII/\HavNII)}$ ratio does not have a significant correlation with stellar mass for CLEAR galaxies.
\label{Fig:unVOM}}
\end{figure*}

\begin{figure*}[t]
\epsscale{1.1}
\plotone{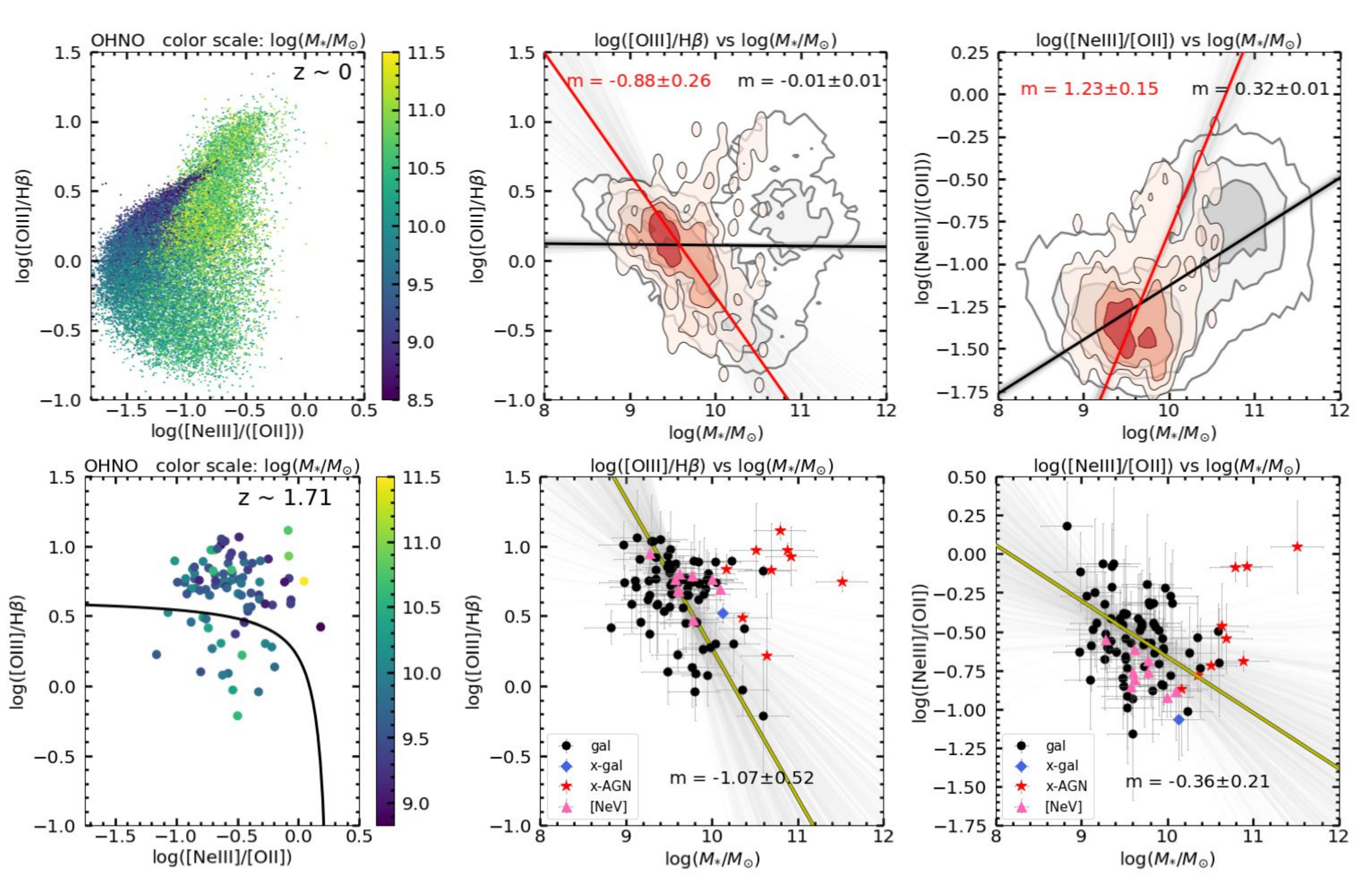}
\figcaption{The relationship between the OHNO emission-line ratios and galaxy stellar mass. The top three panels depict the low-redshift SDSS OHNO galaxies and the bottom three panels depict the high-redshift CLEAR OHNO galaxies. The left panels display the OHNO diagrams with galaxies color-coded by stellar mass, the middle panels display the relation of $\log{(\OIII/\Hb)}$ with stellar mass, and the right panels show the relation between $\log{(\NeIII/\OII)}$ and stellar mass. The middle and right panels include a best-fit linear regression line for the non-AGN (excluding the X-ray AGN and \NeV\ galaxies). The top middle and right panels include two sets of contours and fits: grey/black to the entire SDSS sample and red for a subsample of the SDSS galaxies that are matched to the same stellar mass distribution as the CLEAR sample. As in Figure \ref{Fig:unVOM}, both the CLEAR and mass-matched SDSS galaxies have a significant anti-correlation between $\log{(\OIII/\Hb)}$ and stellar mass. CLEAR galaxies also have a marginal anti-correlation of $\log{(\NeIII/\OII)}$ with stellar mass, although interpreting this is limited by the small sample size and limited dynamic range in stellar mass.
\label{Fig:OHNOM}}
\end{figure*}

Figures \ref{Fig:unVOM} and \ref{Fig:OHNOM} show the relation between the unVO87 and OHNO line ratios with stellar mass. The top three panels show the SDSS galaxies and the bottom row shows the CLEAR sample. The left-most panels show either the unVO87 or OHNO diagrams color-coded by $\log(\Ms)$. The center and right columns show how each emission-line ratio relates to stellar mass, with a best-fit line for non-AGN (excluding the X-ray AGN and \NeV\ galaxies) calculated from the python \texttt{linmix} linear regression package. The top middle and right panels have two best-fit lines and contours. The grey contours represent the entire SDSS sample, with a black best-fit line calculated from a random subset of 2000 SDSS galaxies (in order to reduce computational time for the linear regression fit). The red contours show a subset of the SDSS sample that is matched to have the same distribution of stellar mass as the CLEAR sample, randomly drawing 5 SDSS galaxies from stellar mass bins of 0.1~dex width for every CLEAR galaxy. The red line shows the best linear regression fit to the mass-matched subset of SDSS galaxies.

\begin{figure*}[t!]
\epsscale{1.1}
\plotone{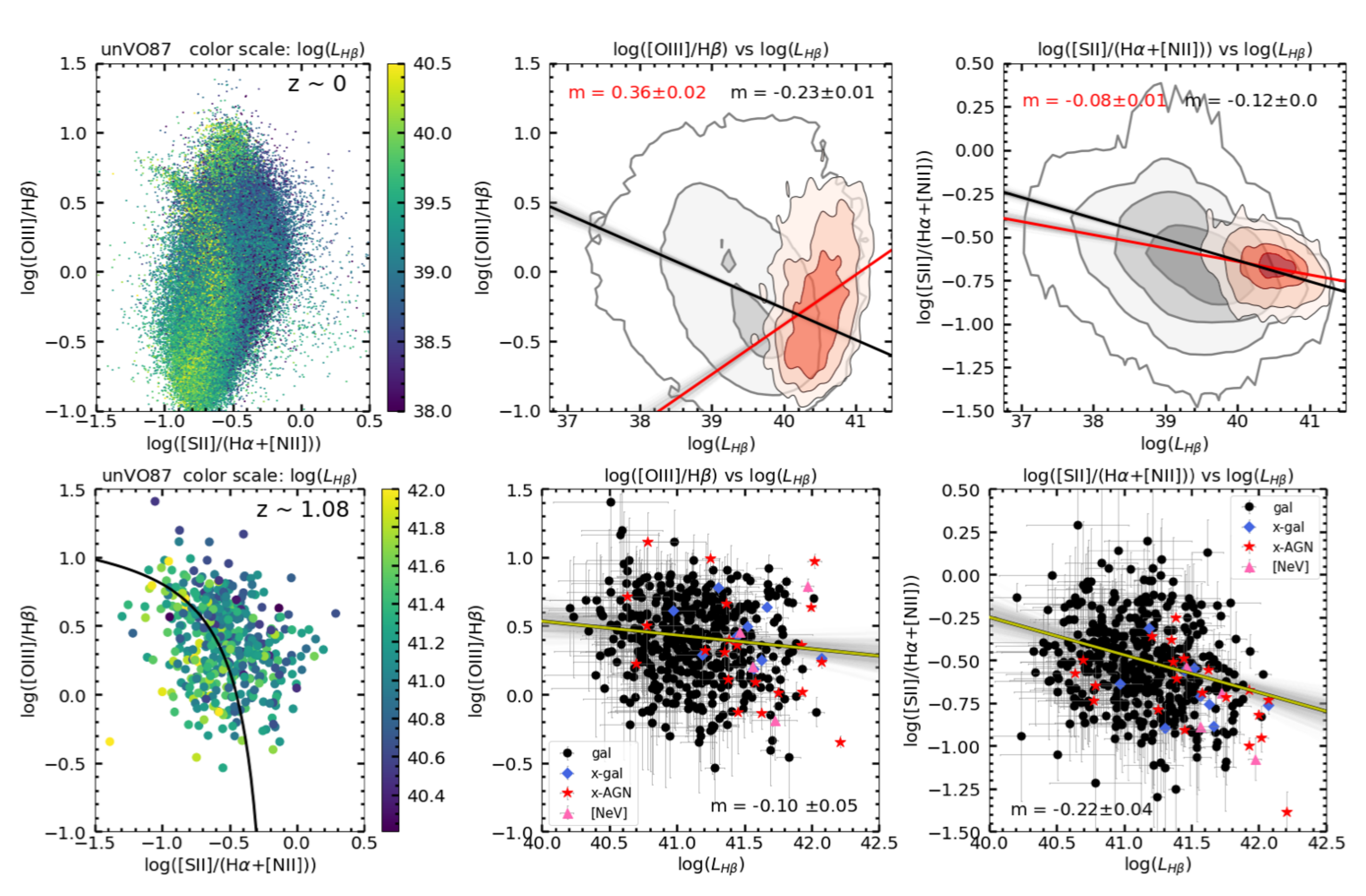}
\figcaption{The relationships between the unVO87 emission-line ratios and the galaxy \Hb\ luminosity. The top three panels depict the low-redshift SDSS unVO87 data and the bottom three panels depict the high-redshift CLEAR unVO87 data. The left panels display the unVO87 diagram with galaxies color-coded by $\log(L_{\Hb})$, the middle panels display the relation of $\log{(\OIII/\Hb)}$ with $\log(L_{\Hb})$, and the right panels show the relation of $\log{(\SII/\HavNII)}$ with $\Hb$ luminosity. The middle and right panels include a best-fit linear regression line for the non-AGN (excluding the X-ray AGN and \NeV\ galaxies). The top middle and right panels include two fits and contours, grey/black to the entire SDSS sample and red for the evolution-matched subsample of the SDSS galaxies with \Hb\ luminosities matching the distribution of CLEAR galaxy \Hb\ luminosities translated to $z=0$ following Equation \ref{eq:zcor}.
There is a significant anti-correlation between $\log{(\SII/\HavNII)}$ and $\log(L_{\Hb})$ for high-redshift CLEAR galaxies. The full sample of low-redshift SDSS galaxies has a significant anti-correlation between $\log{(\OIII/\Hb)}$ and $\log(L_{\Hb})$. The sub-sample does not show this anti-correlation, likely due to its restricted range of $\log(L_{\Hb})$. CLEAR galaxies have a marginal (2$\sigma$) anti-correlation between $\log{(\OIII/\Hb)}$ and $\log(L_{\Hb})$, although we later use multiple linear regression to find that, after accounting for the separate anti-correlation with stellar mass, $\log{(\OIII/\Hb)}$ is significantly correlated with $\log(L_{\Hb})$ (see also Papovich et al. in prep).
\label{Fig:unVOHb}}
\end{figure*}

\begin{figure*}[t!]
\epsscale{1.1}
\plotone{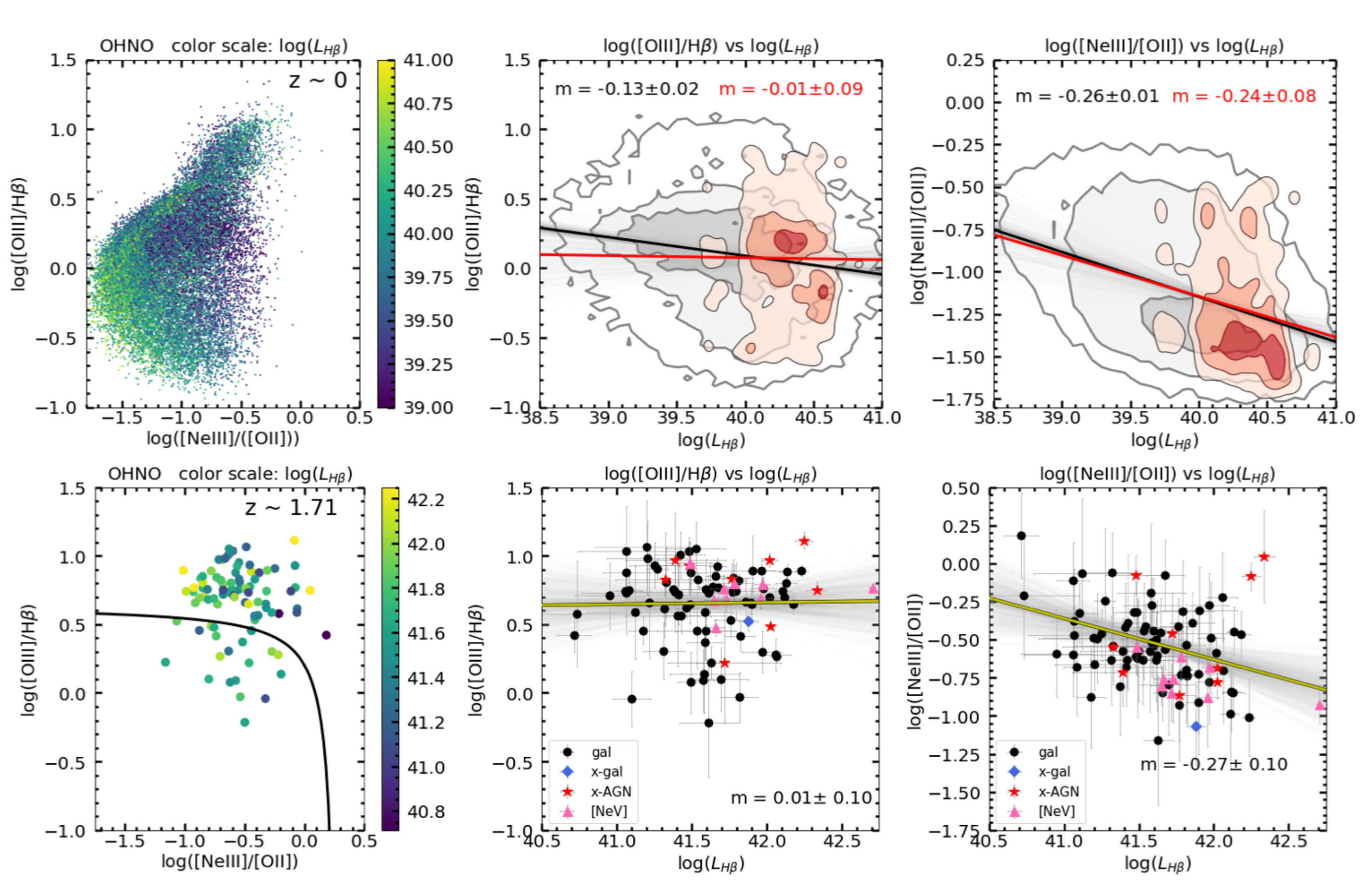}
\figcaption{The relationships between the OHNO emission-line ratios and the galaxy \Hb\ luminosity. The top three panels depict the low-redshift SDSS OHNO data and the bottom three panels depict the high-redshift CLEAR OHNO data. The left panels display the OHNO diagram with galaxies color-coded by $\log(L_{\Hb})$, the middle panels display the relation of $\log{(\OIII/\Hb)}$ with $\log(L_{\Hb})$, and the right panels show the relation between $\log{(\NeIII/\OII)}$ and $\log(L_{\Hb})$. The middle and right panels include a best-fit linear regression line for the non-AGN (excluding the X-ray AGN and \NeV\ galaxies). The top middle and right panels include two sets of contours and fits: grey/black to the entire SDSS sample and red for the evolution-matched subsample of the SDSS galaxies with \Hb\ luminosities matching the distribution of CLEAR galaxy \Hb\ luminosities translated to $z=0$ following Equation \ref{eq:zcor}. There is a marginal anti-correlation between $\log{(\NeIII/\OII)}$ and $\log(L_{\Hb})$ among both CLEAR galaxies and the evolution-matched SDSS sample. There are no apparent correlations between $\log{(\OIII/\Hb)}$ and $\log(L_{\Hb})$, likely in part due to the limited size of the CLEAR OHNO galaxy sample.
\label{Fig:OHNOHb}}
\end{figure*}

The largest difference between the SDSS and CLEAR samples is the lack of massive galaxies at high redshift: the SDSS sample includes galaxies spanning $8<\log(\Ms)<12$, while the CLEAR galaxies are largely limited to $\log(\Ms)<10.5$ with most of the massive CLEAR galaxies with $\log(\Ms)<10.5$ detected as X-ray AGN. There is a statistically significant anti-correlation ($>$3$\sigma$) between $\log(\OIII/\Hb)$ and stellar mass in the unVO87 CLEAR sample and in the unVO87 and OHNO SDSS mass-matched samples. Although the SDSS and CLEAR samples have consistent relationships of $\log(\OIII/\Hb)$ with stellar mass, the high-redshift CLEAR galaxies are shifted to $\log(\OIII/\Hb)$ ratios that are $\sim$0.5 dex higher (as see in Figure \ref{Fig:O3vsz}). \citet{Kash17} similarly showed that $z\sim1.6$ galaxies have higher $\OIII/\Hb$ at fixed stellar mass compared to local galaxies.

The anti-correlation between $\log(\OIII/\Hb)$ and stellar mass has been shown in \citet{dick16} and \citet{Kash19}, and is due to lower metallicity and higher ionization in galaxies with higher sSFR (see Section 5.2). Figure \ref{Fig:unVOM} also shows there is no significant correlations between the $\log(\SII/\HavNII)$ with stellar mass for CLEAR galaxies.
This matches the finding of \citet{Kash17} for low-mass ($\log(\Ms/M_\odot)<10.5$) galaxies. \citet{Kash17} note that massive ($\log(\Ms/M_\odot)>10.5$) galaxies do tend to have lower $\log(\SII/\HavNII)$ ratios than local galaxies of the same mass, but we cannot make a similar comparison in our sample because most of the $\log(\Ms/M_\odot)>10.5$ galaxies are X-ray AGN that are excluded from the fit.

Figure \ref{Fig:OHNOM} shows that CLEAR galaxies have a marginal (2$\sigma$) anti-correlation of $\log(\NeIII/\OII)$ with with stellar mass. The mass-matched SDSS sample has an opposite trend of $\log(\NeIII/\OII)$ increasing with stellar mass. The trend in the SDSS galaxies is likely driven by the handful of high-$\log(\NeIII/\OII)$ at the highest masses: these may be X-ray or \NeV\ AGN that were otherwise rejected from the CLEAR sample.

Figures \ref{Fig:unVOHb} and \ref{Fig:OHNOHb} investigate the relationships between the unVO87 and OHNO emission-line ratios with the $\Hb$ luminosity. We assume that $L(\Hb)$ is a proxy for the unobscured galaxy SFR following Equation \ref{Eq:SFR}. It is possible that higher $L(\Hb)$ results from lower dust rather than higher SFR, but most of the galaxies in our sample are unlikely to have high attenuation due to being emission-line selected, and so any correlations of the line ratios with $L(\Hb)$ are more likely to be caused by differences in SFR than by differences in dust attenuation. As in Figures \ref{Fig:unVOM} and \ref{Fig:OHNOM}, the largest difference between the SDSS and CLEAR samples is in the x-axis: the high-redshift CLEAR galaxies tend to have higher SFR than the low-redshift SDSS galaxies. The red contours represent a subset of the SDSS sample that is matched to have a distribution of \Hb\ luminosities that are ``evolution-matched'' to the \Hb\ luminosities of CLEAR galaxies translated to $z=0$ using Equation \ref{eq:zcor}. The evolution-matched SDSS galaxy sample is constructed by randomly selecting 5 galaxies from $\pm$0.1~dex of the $z=0$ \Hb\ luminosity of each CLEAR galaxy.

The full sample of SDSS galaxies in Figures \ref{Fig:unVOHb} and \ref{Fig:OHNOHb} has significant anti-correlations of all three line ratios with $\log(L_{\Hb})$. These trends generally weaken or disappear entirely when considering the ``evolution-matched'' SDSS galaxies, probably in part due to the much smaller dynamic range in $\log(L_{\Hb})$ of this subsample. The CLEAR galaxies have a significant anti-correlation of $\log(\SII/\HavNII)$ with $\log(L_{\Hb})$ that is consistent with the trend seen among SDSS galaxies. The CLEAR galaxies do not have a significant relationship between $\log(\OIII/\Hb)$ and $\log(L_{\Hb})$, although this is complicated by the inter-correlation of SFR with stellar mass and the previously noted correlation of $\log(\OIII/\Hb)$ with stellar mass. Below we use multiple linear regression to distinguish the strongest relationships of line ratios with the interrelated quantities of galaxy stellar mass, SFR, and redshift. Figure \ref{Fig:OHNOHb} also shows a marginal ($2.7\sigma$) anti-correlation of $\log(\NeIII/\OII)$ with $\log(L_{\Hb})$ among CLEAR galaxies, with a slope that is consistent with the same trend among evolution-matched SDSS galaxies. Again, this trend is complicated by the inter-correlations between SFR and stellar mass.

We use multiple linear regression, implemented by the IDL \texttt{mlinmix\_err} routine \citep{kell07}, to test which of redshift, stellar mass, and SFR are most effectively correlated with the line ratios. Here we assume that the redshift, stellar mass, and $\log(L_{\Hb})$ of each galaxy are measured independently, which is reasonable even though the parameters themselves are physically correlated. The results of the multiple linear regression are given by Table \ref{tab:mlinmix}. This analysis suggests that the most significant correlations of galaxy properties with line ratios are a decrease of $\log(\OIII/\Hb)$ with stellar mass, and an increase of $\log(\OIII/\Hb)$ and decrease of $\log(\SII/\HavNII)$ with $\log(L_{\Hb})$ (or SFR). Interestingly, the multiple linear regression indicates that there is no significant correlation between $\log(\OIII/\Hb)$ and redshift, suggesting that the correlation observed in Section 4 is associated with the decrease in stellar mass with redshift. Our data do not show evidence for mass-independent redshift evolution of $\log(\OIII/\Hb)$ line ratios.

\begin{deluxetable*}{l|rr|rr}[t]
\tablecaption{Multiple Linear Regression Results \label{tab:mlinmix}}
\tablenum{3}
\tablecolumns{5}
\tablewidth{0pt}
\tablehead{\colhead{Parameter} & \multicolumn{4}{c}{Correlation Slopes}}
\startdata
 & \multicolumn{2}{c}{unVO87 ($0.6<z<1.5$)} & \multicolumn{2}{c}{OHNO ($1.2<z<2.5$)} \\
 & \OIII/\Hb & \SII/\HavNII & \OIII/\Hb & \NeIII/\OII \\
\hline
Redshift &  0.13$\pm$0.11 &       0.22$\pm$0.08 &     -0.10$\pm$0.33 &      -0.08$\pm$0.57 \\
$\log(\Ms)$ & \textbf{-0.65$\pm$0.08} &       0.12$\pm$0.06 &     -1.45$\pm$1.10 &      -0.48$\pm$1.76 \\
$\log(L_{\Hb})$ &  \textbf{0.33$\pm$0.10} &       \textbf{-0.36$\pm$0.08} &     1.00$\pm$0.86 &      0.12$\pm$1.66 \\
Excess scatter & 0.14 &       0.17 &     0.14 &      0.09 \\
\enddata
\caption{Results from a multiple linear regression fit for the relationship of the observed line ratios with galaxy redshift, stellar mass, and \Hb\ luminosity. Each cell gives the slope and uncertainty for the related parameter, and the last row gives the excess scatter of the fit. Significant correlations, with slopes that are $>$3$\sigma$ different from zero, are indicated in bold.}
\end{deluxetable*}

Summarizing our investigation of line ratios with galaxy properties:

\begin{itemize}
    \item Both high-redshift CLEAR and SDSS galaxies have $\log(\OIII/\Hb)$ line ratios that are significantly anti-correlated with stellar mass, and positively correlated with $\log(L_{\Hb})$ (or by extension, with $\log(\rm SFR)$).
    
    \item The $\log(\OIII/\Hb)$ line ratio is not correlated with redshift in a way that is independent of stellar mass or SFR.

    \item $\log(\SII/\HavNII)$ is significantly anti-correlated with $\log(L_{\Hb})$ (or by extension, with $\log(\rm SFR)$) in both CLEAR and SDSS galaxies.

    \item CLEAR galaxies have marginal anti-correlations of $\log(\NeIII/\OII)$ with stellar mass and $\log(L_{\Hb})$. The strength of these correlations is likely limited by the small sample size of our study, and improved understanding will require larger samples of high-redshift galaxies: for example, made available from the \textit{James Webb Space Telescope} (JWST).

    \item Interpreting the correlations of line ratios with individual galaxy properties is complicated by the known inter-correlations between redshift, stellar mass, and SFR (e.g. \citealp{mada14}). We use multiple linear regression to disentangle these inter-correlations and find somewhat different relationships than are apparent from relating line ratios to single galaxy parameters.
\end{itemize}

\subsection{Comparison to Photoionization Models}

\begin{figure}[t]
\epsscale{1.1}
\plotone{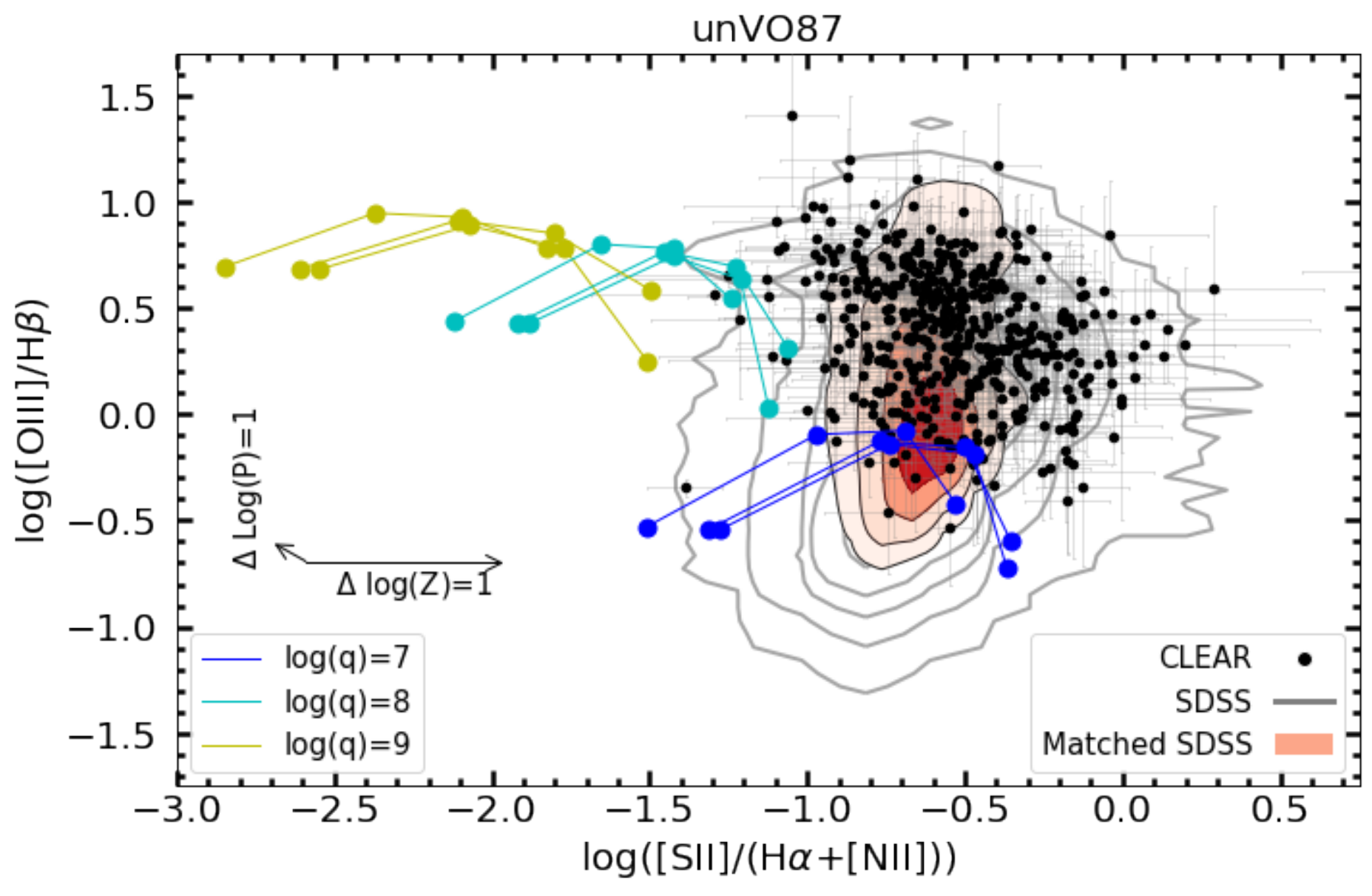}
\figcaption{Theoretical unVO87 line ratio predictions from the MAPPINGS~V models (colored lines) compared to the low-redshift SDSS galaxies (gray contours), evolution-matched SDSS galaxies (red contours) and high-redshift CLEAR galaxies (black points). Color indicates the model ionization and each connected line goes from low to high metallicity ($Z/Z_\odot = 0.05$ to $Z/Z_\odot = 1.0$). Model lines are repeated for three values of pressures. Inset vectors indicate the direction and amplitude of 1~dex increases in metallicity and pressure. The main loci of both the evolution-matched and full SDSS samples are well-described by models with low ionization and moderate metallicity. But the theoretical models do not overlap with the high-redshift CLEAR galaxies or with the tail of the SDSS galaxies, likely because the plane-parallel geometry of the MAPPINGS~V models do not accurately describe the detailed geometry of the \SII\ emission \citep{xiao18}.
\label{Fig:unVOmap}}
\end{figure}

\begin{figure}[t]
\epsscale{1.1}
\plotone{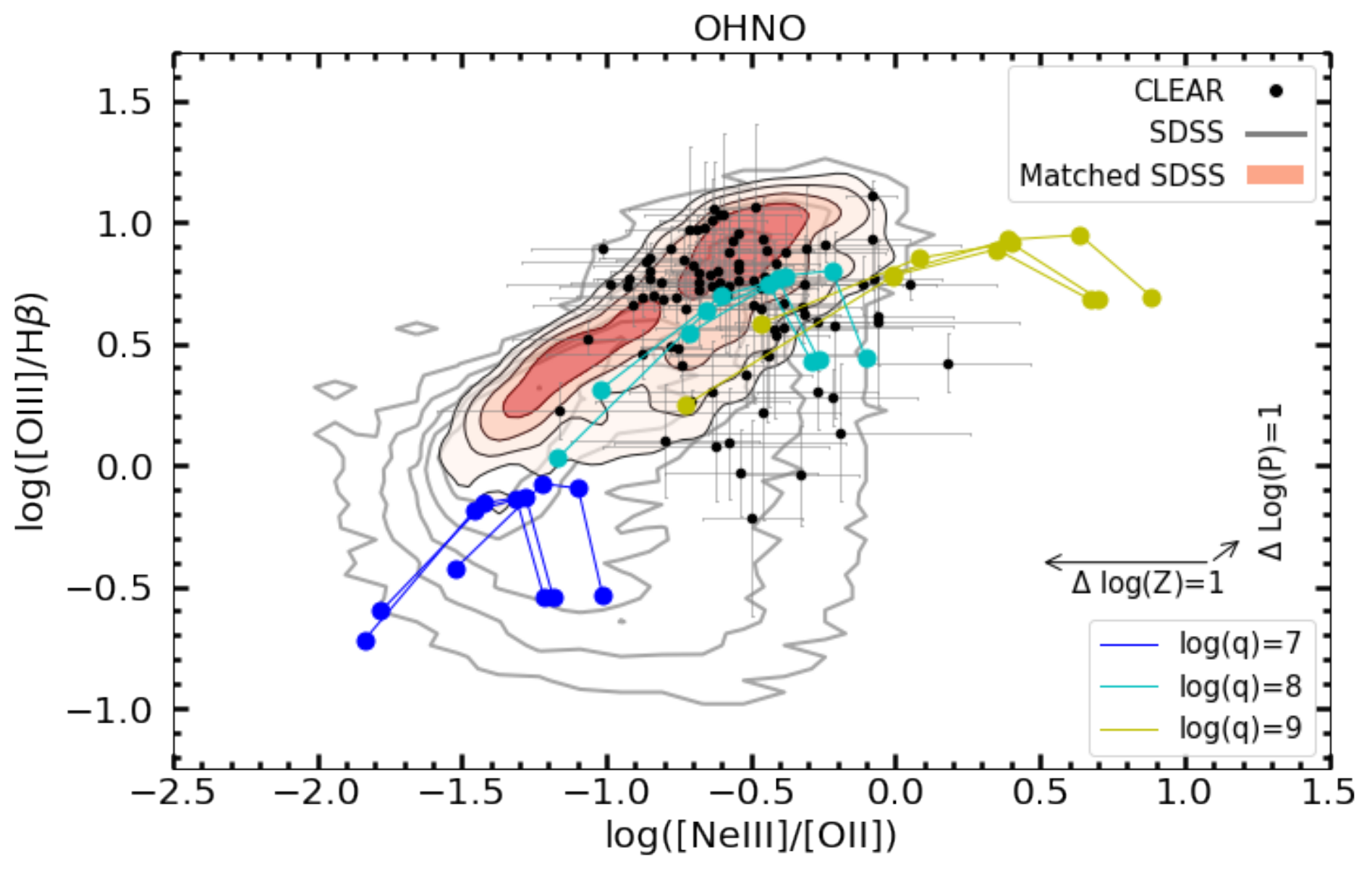}
\figcaption{Theoretical unVO87 line ratio predictions from the MAPPINGS~V models (colored lines) compared to the low-redshift SDSS galaxies (gray contours), evolution-matched SDSS galaxies (red contours) and high-redshift CLEAR galaxies (black points). Model ionization is indicated by the color of the line, and metallicity decreases from left to right for each set of connected model points. Lines for three different pressures are shown, but as in Figure \ref{Fig:unVOmap} the line ratios are not strongly dependent on pressure. Inset vectors indicate the direction and amplitude of 1~dex increases in metallicity and pressure. The evolution-matched and full SDSS samples are well-described by models with low to moderate ionization and low to moderate metallicity, while the CLEAR galaxies are better described by models with higher ionization and lower metallicity.
\label{Fig:OHNOmap}}
\end{figure}

We compare our measured emission-line ratios with theoretical models to determine the physical conditions that are likely to be responsible for the observed correlations of the line ratios with stellar mass and  $\log(L_{\Hb})$. We use the emission-line model of \citet{Kewl19}, which combines Starburst99 \citep{Leit99} models of stellar ionizing spectra with the MAPPINGS~V photoionization code \citep{Suth18}. The Starburst99 model spectra use a \citet{salp55} IMF and include mass loss. The MAPPINGS~V code uses atomic data from the CHIANTI~8 database \citep{dere97,delz15} and includes photoionization, recombination, excitation, and dust depletion in the model \HII\ regions. We use the "Pressure Models" of \citet{Kewl19} that create synthetic emission-line spectra over a grid of pressure $\log(P/k)$, ionization $\log(q)$, and metallicity $Z/Z_\odot$, interpolating between Starburst99 models and CHIANTI~8 data as needed to match the grid. We use synthetic emission line spectra with the following values of pressure, ionization, and metallicity:
\begin{itemize}
    \item Pressure $\log(P/k) = [6,7,8]$, units of cm$^{-3}$
    \item Ionization $\log(q) = [7,8,9]$, units of cm~s$^{-1}$
    \item Metallicity $Z/Z_\odot = [0.05,0.2,0.4,1.0]$
\end{itemize}
The lowest metallicity bin is not well-constrained by observations and is extrapolated in the Starburst99 input spectra, and so the $Z/Z_\odot = 0.05$ synthetic spectra are likely to be the least certain of the theoretical predictions.

Figures \ref{Fig:unVOmap} and \ref{Fig:OHNOmap} compare the observed unVO87 and OHNO line ratios with the MAPPINGS~V theoretical predictions for different pressure, ionization, and density. Each colored line represents models with fixed pressure and ionization and four different metallicities ($Z/Z_\odot = [0.05,0.2,0.4,1.0]$), with the color of the line corresponding to the ionization. Lines of each color (and ionization) are shown for the three values of $\log(P/k)$. Vectors inset in the plot also show the typical scale and direction of $\sim$1~dex changes in metallicity and pressure. Metallicity increases from left to right in the unVO87 diagram and decreases from left to right in the OHNO diagram. The $\OIII/\Hb$ ratio increases with ionization and is double-valued at high and very low metallicities. All of the line ratios are relatively insensitive to differences in pressure (the $\log(P/k)=6$ and $\log(P/k)=7$ lines are nearly identical).

The bulk of low-redshift SDSS galaxies and evolution matched SDSS galaxies, shown by gray and red contours in Figures \ref{Fig:unVOmap} and \ref{Fig:OHNOmap}, are well-described by theoretical line ratios produced in \HII\ regions with moderate to low ionization and moderate to high metallicity. The OHNO line ratios of the high-redshift CLEAR galaxies are similarly well-described by theoretical spectra from high ionization and low metallicity. On the other hand, the MAPPINGS~V models do not effectively describe the unVO87 line ratios of the CLEAR galaxies (along with the tail of the distribution of low-redshift SDSS galaxies). Theoretical predictions of \SII\ emission lines are particularly sensitive to the geometry of the \HII\ region \citep{xiao18}, and these details are likely to be missing in the idealized plane-parallel geometry of the MAPPINGS~V models. In particular, the MAPPINGS-V spectra are meant to simulate \HII\ regions, but the diffuse gas between \HII\ regions is also likely to emit a significant amount of the \SII\ that is observed in an integrated galaxy spectrum.

In Section 5.1, we found that the most significant line-ratio correlations were increasing $\OIII/\Hb$ with decreasing stellar mass and increasing SFR, and increasing $\SII/\HavNII$ with decreasing SFR. We can use the theoretical MAPPINGS~V predictions to connect these line-ratio trends to the physical conditions of the gas. The connection between $\OIII/\Hb$ and stellar mass is likely caused by decreasing metallicity in low-mass galaxies, as has been observed in the canonical mass-metallicity relation \citep[e.g.,][]{trem04}. The relationships of $\OIII/\Hb$ and $\SII/\HavNII$ with $L(\Hb)$ is likely caused by increased ionization associated with higher SFR (\citealt{Brin08, Liu08}, Papovich et al. in prep.), with $\OIII/\Hb$ increasing and $\SII/\HavNII$ decreasing at higher ionization. \cite{Kash17} similarly found that higher ionization of the ISM would lead to lower $\SII/\Ha$ and higher $\OIII/\Hb$ compared to local galaxies. The relationships of $\OIII/\Hb$ and $\SII/\HavNII$ with stellar mass and $L(\Hb)$ match previous measurements of the relationship between metallicity, stellar mass, and star formation rate \citep[e.g.,][]{lilly13, salim14, wuyt14, zahi14}.

The MAPPINGS~V models show that $\NeIII/\OII$ increases at higher ionization and lower metallicity. We would then expect to see correlations of increasing $\NeIII/\OII$ with decreasing stellar mass and increasing SFR, analogous to the observed correlations for $\OIII/\Hb$. The lack of significant correlations for $\NeIII/\OII$ may be due to the limitations of the sample. Our CLEAR OHNO sample includes only 91 galaxies (and 73 non-AGN), and is limited to narrow ranges of (low) stellar mass and (high) $L(\Hb)$. Most of the non-AGN $\NeIII/\OII$ line ratios are just above the detection limit (e.g., Figure \ref{Fig:Ne3vsz}), and so may represent only part of a larger distribution of $\NeIII/\OII$ among the broader galaxy population. Our work implies that $\NeIII/\OII$ and the OHNO diagram are an effective indicator of AGN detected in X-rays or by $\NeV$. But understanding the relationships of $\NeIII/\OII$ with galaxy physical conditions in non-AGN would require a larger sample with deeper line flux limits: for example, from future spectroscopic surveys with the \textit{James Webb Space Telescope}.

\section{Summary} 

We have defined the unVO87 diagram using 461 galaxies at $z \sim 1.08$ and OHNO diagram using 91 galaxies at $z \sim 1.69$ using the CLEAR survey observed by \textit{HST}. The sample is selected by having $\mathrm{SNR} > 1$ in all emission lines and is visually inspected to remove any contaminated galaxies. We have defined the AGN/SF line for the diagram unVO87 by using the known AGN/SF line of the VO87 diagram with a correction for the blending of our $\Ha$ line with $\NII$. The OHNO AGN/SF line is defined by taking galaxies that are in both the unVO87 and OHNO diagrams and empirically choosing a line that separates unVO87 AGN-classified galaxies from the unVO87 SF-classified galaxies.

We studied the effectiveness of the unVO87 and OHNO diagrams in several aspects and summarize our results as follows:

\begin{itemize}

\item The OHNO diagram effectively separates X-ray AGN and \NeV\ sources from the remaining galaxy population at $z>1$. Although the unVO87 diagram is effective at $z \sim 0$, it does not effectively distinguish AGN and non-AGN galaxies at higher redshift. To show this further we preformed statistical tests on the OHNO diagram, demonstrating that the AGN and \NeV\ galaxies have a different parent distribution in the $\NeIII/\OII$ line ratio than non-AGN galaxies (Figures \ref{fig:unVO87} and \ref{fig:OHNO}).

\item We compare ``evolution-matched'' samples of low-redshift SDSS and high-redshift CLEAR galaxies, matched to have the same limiting luminosity while controlling for evolution of the SFR-mass diagram, showing that $z \sim 1$ galaxies have $\sim$0.1~dex higher $\SII/\HavNII$ and $\sim$0.5~dex higher $\OIII/\Hb$ than $z \sim 0$ galaxies (Figures \ref{Fig:unvo87levo} and \ref{Fig:unvo87hevo}).

\item Comparing ``evolution-matched'' samples of SDSS and CLEAR galaxies, we find that $z \sim 1.7$ galaxies have $\sim$0.2~dex higher $\NeIII/\OII$ than $z \sim 0$ galaxies. CLEAR galaxies have much higher average ratios of $\NeIII$ to $\OIII$, despite the similar ionization potentials of the two lines, and there are no SDSS galaxies with $\NeIII$ as luminous as found in the high-redshift CLEAR galaxies (Figures \ref{Fig:OHNOlevo}-\ref{Fig:Ne3vsO3}.)

\item Multiple linear regression indicates that the $\OIII/\Hb$ line ratio is significantly anti-correlated with stellar mass (Figures \ref{Fig:unVOM} and \ref{Fig:OHNOM}). We also find significant correlations of increasing $\OIII/\Hb$ and decreasing $\SII/\HavNII$ with with $L_{H_\beta}$, assumed to be a proxy for the unobscured SFR of a galaxy (Figures \ref{Fig:unVOHb} and \ref{Fig:OHNOHb}). MAPPINGS-V models indicate that these relationships are due to lower metallicity and higher ionization in galaxies of lower stellar mass and higher SFR (Figures \ref{Fig:unVOmap} and \ref{Fig:OHNOmap}).

\item Low-redshift SDSS galaxies are well-described by MAPPINGS-V model spectra with moderate/low ionization and moderate/high metallicity. The CLEAR OHNO line ratios are similarly well-described by model spectra with high ionization and low metallicity. The unVO87 line ratios of CLEAR galaxies, on the other hand, are not well-described by MAPPINGS-V, likely due to $\SII$ emission emerging from diffuse gas that is not included in the idealized $\HII$ region geometry of the models (Figures \ref{Fig:unVOmap} and \ref{Fig:OHNOmap}).

\item Although we find that the $\NeIII/\OII$ line ratio and OHNO diagram are effective at separating X-ray and $\NeV$ AGN from the rest of the galaxy population, we do not find any significant correlations of $\NeIII/\OII$ with stellar mass or SFR in non-AGN galaxies. This is likely due to the limited size, dynamic range, and line flux limit of our high-redshift CLEAR sample.

\end{itemize}

Our results are useful to consider for upcoming spectroscopic surveys with the \textit{James Webb Space Telescope} (\textit{JWST}). In particular, we find that the OHNO diagram of $\NeIII/\OII$ versus $\OIII/\Hb$ is effective for distinguishing AGN from other $z>1$ galaxies with lower ionization, while the unVO87 diagram (a low-resolution version of the classic \citealt{veil87} VO87 diagnostic) becomes ineffective at high redshift. We also show that the $\OIII/\Hb$ and $\SII/\HavNII$ line ratios indicate lower metallicity and higher ionization in galaxies with lower stellar mass and higher SFR, with no redshift evolution in the line ratios beyond the trends with mass and SFR. \textit{JWST} near-IR spectroscopy will be able to observe these line ratios over $2<z<8$, unraveling the AGN content and physical conditions of galaxies from the peak of cosmic SFR to the first galaxies at cosmic dawn.

\section{Acknowledgements}
This work is based on data obtained from the Hubble Space Telescope through program number GO-14227.  Support for program GO-14227 was provided by NASA through a grant from the Space Telescope Science Institute, which is operated by the Association of Universities for Research in Astronomy, Incorporated, under NASA contract NAS5-26555. BEB, JRT, and NJC acknowledge support from NSF grant CAREER-1945546 and NASA grant JWST-ERS-01345.
RCS appreciates support from a Giacconi Fellowship at the Space Telescope Science Institute. VEC acknowledges support from the NASA Headquarters under the Future Investigators in NASA Earth and Space Science and Technology (FINESST) award 19-ASTRO19-0122, as well as support from the Hagler Institute for Advanced Study at Texas A\&M University. This work benefited from generous
support from the George P. and Cynthia Woods Mitchell Institute for
Fundamental Physics and Astronomy at Texas A\&M University. 
\bibliography{lib}{}

\end{document}